\documentclass[12pt]{article}
\usepackage{graphicx} 
\def\laq{\raise 0.4ex\hbox{$<$}\kern -0.8em\lower 0.62
ex\hbox{$\sim$}}
\def\gaq{\raise 0.4ex\hbox{$>$}\kern -0.7em\lower 0.62
ex\hbox{$\sim$}}

\begin{document}

\begin{titlepage}
\begin{flushright}
CERN-PH-TH/2008-021
\end{flushright}
\vspace*{1cm}

\begin{center}

{\large{\bf Generalized CMB initial conditions with pre-equality magnetic fields}}
\vskip1.cm

Massimo Giovannini$^{a,c}$ and Kerstin E. Kunze$^{b,c}$

\vskip2.cm

{\sl $^a$Centro ``Enrico Fermi", Compendio del Viminale, Via 
Panisperna 89/A, 00184 Rome, Italy}
\vskip 0.2cm 
{\sl $^b$ Departamento de F\'\i sica Fundamental, \\
 Universidad de Salamanca,Plaza de la Merced s/n, E-37008 Salamanca, Spain}
\vskip 0.2cm 
{\sl $^c$  Department of Physics, Theory Division, CERN, 1211 Geneva 23, Switzerland}

\vspace*{1cm}

\begin{abstract}
The most general initial conditions of CMB anisotropies, compatible with the presence of  pre-equality magnetic fields, are derived.  When the plasma is composed by photons, baryons, electrons, CDM particles and neutrinos,
the initial data of the truncated Einstein-Boltzmann hierarchy contemplate one magnetized adiabatic mode and four (magnetized) non-adiabatic modes. After obtaining the analytical  form of the various solutions, the Einstein-Boltzmann hierarchy is numerically integrated for the corresponding sets of initial data. The TT, TE and EE angular power spectra are illustrated and discussed for the magnetized generalization of the CDM-radiation mode, of the baryon-radiation mode and  of the non-adiabatic mode of the neutrino sector. 
Mixtures of initial conditions are 
examined by requiring that the magnetized adiabatic mode dominates 
over the remaining non-adiabatic contributions. In the latter case,  possible degeneracies between complementary sets of initial data might be avoided through the combined analysis of the TT, TE and EE angular power spectra at high multipoles (i.e. $\ell >1000$).
\end{abstract}

\end{center}
\end{titlepage}

\newpage
\renewcommand{\theequation}{1.\arabic{equation}}
\setcounter{equation}{0}
\section{Pre-equality CMB initial conditions}
\label{sec1}
What are the initial conditions of CMB anisotropies? 
In the pivotal $\Lambda$CDM paradigm, the 
initial conditions are taken to be adiabatic 
\cite{WMAP1,WMAP2,WMAP3,WMAP4,WMAP5}. 
In this case the fluctuations of the total pressure are  proportional, prior to matter-radiation  equality, to the fluctuations of the (total) energy density.
Already in the absence of large-scale magnetic fields, the initial data 
of CMB anisotropies are not exhausted by adiabatic solution.
In the non-adiabatic case the fluctuations of the total pressure arise 
because of the compositeness of the pre-equality plasma. 

Standard Boltzmann solvers include 
the possibility of having one (or more) non-adiabatic initial conditions. 
The generalized initial conditions 
of the Einstein-Boltzmann hierarchy\footnote{By Einstein-Boltzmann 
hierarchy we mean the set of kinetic equations written in curved space and 
supplemented by the contribution of the gravitational inhomogeneities 
obeying the perturbed Einstein equations.} are hereby derived when large-scale magnetic fields are consistently included in the pre-equality plasma. 
The obtained results complement and extend former analyses 
only centered around the magnetized adiabatic mode. 

A stochastic background of large-scale magnetic field can be naturally incorporated in  the physics of the adiabatic initial conditions. The TT angular power spectra\footnote{Following a consolidated terminology,  the angular power spectra of the 
temperature autocorrelations will be denoted, with stenographic 
notation, by TT.  In analog terms EE and TE denote, respectively, the angular power spectra
of the polarization autocorrelations and of the temperature-polarization cross-correlations.}
have been analytically estimated, within the tight-coupling approximation, in \cite{mg1,mg2,mg3}. 
The results of \cite{mg2,mg3} can be used for simplified estimates  of the TT angular power spectra, and, partially \cite{mg1},  of the TE and EE correlations.  A full numerical approach  is however required to confront 
the present \cite{WMAP4,WMAP5} and forthcoming \cite{planck} experimental data. 
This problem was successfully addressed and solved in \cite{giokunz1,giokunz2}. 
Semi-analytical results (obtained via a different treatment of recombination and diffusive effects) agree
with the full numerical calculation: the shape of the
correlated distortions of the first three acoustic peaks is correctly captured by the analytical discussion even if the numerical approach is intrinsically more accurate especially at high multipoles. This 
occurrence strengthen the consistency of the numerical approach and allows 
interesting analytical cross-checks.
In \cite{giokunz1,giokunz2} the only initial conditions examined were 
the ones associated with the magnetized adiabatic mode. 
This choice is prompted by the best 
fit of the WMAP data alone as well as, for instance, by the best fits of the WMAP data combined with the large-scale structure data \cite{LSS1,LSS2}, with the supernova data \cite{SN1,SN2} and with all cosmological data sets. The 
strict adiabaticity of the initial conditions will be now relaxed
and  non-adiabatic modes will be scrutinized in the presence of stochastic magnetic fields. This program is technically
mandatory and physically relevant.

If the initial data of the Einstein-Boltzmann hierarchy are solely non-adiabatic, the measured TE correlations cannot be reproduced \cite{WMAP3}. This result is already apparent from the first $200$ multipoles of the TE power spectra where, generically, non-adiabatic modes lead to a positive correlation while a predominant adiabatic mode would imply, instead, a negative cross-correlation. The position of the anticorrelation peak can be related, in the adiabatic case, to the position of the first 
Doppler peak of the TT power spectra (i.e. $\ell_{\mathrm{Dop}} \simeq 220$). The position 
of the first anticorrelation peak of the TE angular power spectrum can be estimated as $\ell_{\mathrm{anti}} \simeq 
3 \ell_{\mathrm{Dop}}/4 \simeq 150$ to first-order in the well known tight-coupling expansion \cite{peeyu} (see also \cite{harzal,kot,sel}). 
The experimental evidence does not exclude that a predominant adiabatic mode could be present in combination with sub-dominant non-adiabatic contributions so that the overall fit to the data may even improve \cite{H4,H5}. 

If the magnetic fields are present prior to matter-radiation equality (and, a fortiori, 
prior to photon decoupling) the physical situation changes when compared with 
the conventional non-adiabatic initial conditions. Plausible questions 
then arise:
\begin{itemize}
\item{} can magnetic field be physically compatible with the various non-adiabatic initial conditions?
\item{} in the latter cases, what are the initial conditions to be imposed to the Einstein Boltzmann hierarchy?
\item{} how do the magnetized non-adiabatic mode affect the TT, EE and TE angular power spectra?
\item{} is it possible to compensate the distortions induced by the magnetized adiabatic mode with an appropriate non-adiabatic contribution where the magnetic field is also consistently included?
\end{itemize}
In the aforementioned list of items, the first pair of questions  
implies a thorough analytical discussion of the initial conditions 
of the Einstein-Boltzmann hierarchy. The remaining two points 
necessarily demand a full numerical integration which will be performed 
with the code devised in \cite{giokunz1,giokunz2} and 
hereby extended to handle initial data which are, simultaneously, magnetized 
and non-adiabatic. The code employed for the numerical integration is based, originally,
on CMBFAST \footnote{In CMBFAST 
only the CDM-radiation and baryon-radiation modes are included. 
In our code all entropic modes are be implemented 
in combination with a stochastic magnetic field affecting both the initial conditions and the 
evolution equations.} \cite{cmbfast1,cmbfast2} (which is, in turn, deeply rooted on COSMICS \cite{cosmics1,cosmics2}). 

To commence,  non-adiabatic
 initial conditions of the Einstein-Boltzmann hierarchy will be 
 quantitatively introduced.  Prior to equality, the plasma consists of a baryon-lepton component, tightly 
coupled via Coulomb scattering, and  supplemented by photons, neutrinos and cold dark matter particles (CDM in what follows). 
The fluctuations of the total pressure $p_{\mathrm{t}}$ (denoted as 
$\delta_{\mathrm{s}} p_{\mathrm{t}}$) can be written, in general terms,  as the sum of two physically different contributions, namely,
\begin{equation}
\delta_{\mathrm{s}} p_{\mathrm{t}} = c_{\mathrm{st}}^2 \delta_{\mathrm{s}} \rho_{\mathrm{t}} + \delta_{\mathrm{s}} p_{\mathrm{nad}}, \hspace{1cm} c_{\mathrm{st}}^2 = \frac{p_{\mathrm{t}}'}{\rho_{\mathrm{t}}'},
\label{deltap}
\end{equation}
where $\delta_{\mathrm{s}} \rho_{\mathrm{t}}$ is the fluctuation of the energy density and 
$c_{\mathrm{st}}^2$ is the total sound speed across the radiation-matter transition\footnote{ The  background geometry  is characterized by a conformally flat line element $ds^2 = a^2(\tau)[d\tau^2 - d\vec{x}^2]$. The scale 
factor $a(\tau)$ will be often referred to its value at matter-radiation 
equality, i.e. $\alpha = a/a_{\mathrm{eq}}$.
Throughout the paper the prime denotes
a derivative with respect to the conformal time coordinate $\tau$ while the overdot indicates a derivative with respect to the cosmic time coordinate $t$. The conformal time coordinate $\tau$ is related to the cosmic time as $a(\tau) d\tau = dt $.}. 
At the right hand side of Eq. (\ref{deltap}), the first 
term (i.e. $c_{\mathrm{st}}^2 \delta_{\mathrm{s}} \rho_{\mathrm{t}}$) parametrizes  the adiabatic contribution. 
The second term at the right hand side of Eq. (\ref{deltap}), (i.e. 
$\delta_{\mathrm{s}} p_{\mathrm{nad}}$), accounts for the non-adiabatic 
pressure fluctuations
which can be also written as\footnote{In the present notations  ${\mathcal H} = a'/a = \alpha'/\alpha = a H$ where, by definition,
$H = \dot{a}/a$ is the Hubble rate in cosmic time.}
\begin{equation}
\delta_{\mathrm{s}} p_{\mathrm{nad}} = \sum_{\mathrm{ij}}\frac{\partial p_{\mathrm{t}}}{\partial \varsigma_{\mathrm{ij}}} \delta \varsigma_{\mathrm{ij}} 
 = \frac{1}{6 {\mathcal H} \rho_{\mathrm{t}}'} \sum_{\mathrm{ij}} \rho_{\mathrm{i}}' \rho_{\mathrm{j}}' 
(c_{\mathrm{si}}^2 - c_{\mathrm{sj}}^2) {\mathcal S}_{\mathrm{ij}},\qquad 
{\mathcal S}_{\mathrm{ij}} = \frac{\delta \varsigma_{\mathrm{ij}}}{\varsigma_{\mathrm{ij}}},
\label{DPNAD1}
\end{equation}
where the indices i and j are not tensor indices but denote 
two generic species of the pre-equality plasma.
Furthermore, in Eq. (\ref{DPNAD1}), $c_{\mathrm{si}}^2$ and $c_{\mathrm{sj}}^2$ are the sound speeds of two (generic) species 
of the plasma; $\delta \varsigma_{\mathrm{ij}}$ is the fluctuation of the specific entropy computed for a given pair of species and 
${\mathcal S}_{\mathrm{ij}}$, as indicated, is the relative fluctuation 
of $\varsigma_{\mathrm{ij}}$. By definition, ${\mathcal S}_{\mathrm{ij}} = - {\mathcal S}_{\mathrm{ji}}$: a factor $2$ (included in the denominator at the right hand side of Eq. (\ref{DPNAD1})) avoids 
double counting of entropy modes when the sum extends over all the species 
present in the plasma.

Neglecting for a moment neutrinos as well as baryons and electrons, we can imagine 
a plasma dominated by photons  but with a subleading contribution of CDM particles (characterized by a concentration $\overline{n}$). In this toy example the 
specific entropy $\varsigma \simeq T_{\gamma}^3/\overline{n}$ will then imply $ {\mathcal S}= 3\delta_{\gamma}/4 - \delta_{\mathrm{c}}$ 
(where $\delta_{\gamma}$ and $\delta_{\mathrm{c}}$ are, respectively, the radiation and the CDM density 
contrasts). For the latter estimate, it is useful to stress that $\rho_{\gamma} \propto T_{\gamma}^4$ while $\rho_{\mathrm{c}} \propto \overline{n}$. This sort of heuristic arguments can be phrased in fully gauge-invariant terms with the result that\footnote{The variables $\zeta_{\mathrm{i}}$ and $\zeta_{\mathrm{j}}$ are 
directly expressed in the synchronous coordinate system (see, for 
instance, Eq. (\ref{metric}) of Appendix \ref{APPA}).}
\begin{equation}
{\mathcal S}_{\mathrm{ij}} = \frac{\delta\varsigma_{\mathrm{ij}}}{\varsigma_{\mathrm{ij}}}= - 3 ( \zeta_{\mathrm{i}} - \zeta_{\mathrm{j}}), \qquad \zeta_{\mathrm{i}} = \xi + \frac{\delta_{\mathrm{i}}}{3(w_{\mathrm{i}} + 1)},
\label{DPNAD2}
\end{equation}
where $w_{\mathrm{i}}$ is the barotropic index of a generic species. 
The entropy fluctuations of Eq. (\ref{DPNAD2}) are expressed 
in terms of the $\zeta_{\mathrm{i}}$ and $\zeta_{\mathrm{j}}$ which are,
themselves,  gauge-invariant 
and which become, in the uniform curvature gauge, the density contrasts 
of the single species\footnote{In equally correct terms we could also argue 
that $\zeta_{\mathrm{i}}$ measures the curvature perturbations 
on the hypersurfaces where the energy density of a given species is uniform. Indeed, in the synchronous gauge
$\zeta$ is proportional to $\xi$ which is related to ${\mathcal R}$ (see the following section) which 
is the curvature perturbation on comoving orthogonal hypersurfaces \cite{KS}.}
 \cite{KS} (see also \cite{hw1,hw2}). The gauge-invariant definition of entropy fluctuations 
goes back to the seminal contributions of \cite{KS} and it is commonly employed in the 
generalized discussion of the evolution equations of curvature and entropy fluctuations (see also 
\cite{entr1,entr2} and references therein).  The quantity $\zeta_{\mathrm{i}}$  of Eq. (\ref{DPNAD2}) is expressed 
in the synchronous gauge which will be consistently used in the analytical calculations and in the 
numerical implementation of the code. In Appendix \ref{APPA}  the truncated Einstein-Boltzmann hierarchy is summarized in  the language  of the synchronous gauge. At early times (and in the tight-coupling limit) the truncated Einstein-Boltzmann hierarchy will be used to deduce the magnetized initial conditions for the various modes.  In the pre-equality
plasma we have four different species and, consequently, 
we will have that:
\begin{equation}
\zeta_{\gamma} = \xi + \frac{\delta_{\gamma}}{4},\qquad \zeta_{\nu} = \xi + \frac{\delta_{\nu}}{4},
\qquad \zeta_{\mathrm{c}} = \xi + \frac{\delta_{\mathrm{c}}}{3}, \qquad 
\zeta_{\mathrm{b}} = \xi + \frac{\delta_{\mathrm{b}}}{3},
\label{DPNAD3}
\end{equation}
where $\delta_{\gamma}$, $\delta_{\nu}$, $\delta_{\mathrm{b}}$ and 
$\delta_{\mathrm{c}}$ are the density contrasts of the corresponding 
component evaluated in the synchronous coordinate system. The electrons are tightly coupled to baryons 
through Coulomb scattering and can be treated, for the purpose of the initial conditions, 
as a single species with approximate common temperature \cite{giokunz2}.
The initial conditions of the Boltzmann hierarchy 
are set after neutrino decoupling  (i.e. well before matter-radiation equality). At this epoch, Eqs. (\ref{DPNAD1}), (\ref{DPNAD2})  and (\ref{DPNAD3}) imply that ${\mathcal S}_{\mathrm{cr}}$ is:
\begin{equation}
{\mathcal S}_{\mathrm{c}\gamma} = - 3 ( \zeta_{\mathrm{c}} - \zeta_{\gamma}) = \frac{3}{4} \delta_{\gamma} - \delta_{\mathrm{c}},\qquad {\mathcal S}_{\mathrm{c}\nu} = - 3 ( \zeta_{\mathrm{c}} - \zeta_{\nu}) = \frac{3}{4} \delta_{\nu} - \delta_{\mathrm{c}}.
\label{NAD1}
\end{equation}
Equation 
(\ref{NAD1}) defines the CDM-radiation mode.
The baryon-radiation mode causes, instead, a mismatch between 
the density contrast of the baryon-lepton fluid and the 
density contrasts of the (two) relativistic species at the corresponding 
epoch:
\begin{equation}
{\mathcal S}_{\mathrm{b}\gamma} = - 3(\zeta_{\mathrm{b}} - \zeta_{\gamma}) = \frac{3}{4} \delta_{\gamma} - \delta_{\mathrm{b}}, \qquad {\mathcal S}_{\mathrm{b}\nu} = - 3(\zeta_{\mathrm{b}} - \zeta_{\nu}) = \frac{3}{4} \delta_{\nu} - \delta_{\mathrm{b}}.
\label{NAD2}
\end{equation}
In the $\Lambda$CDM choice of cosmological parameters\footnote{We shall denote, in accordance with the 
established conventions, $\omega_{\mathrm{X}0} =h_{0}^2 \Omega_{\mathrm{X}0}$ where 
$h_{0}$ is the indetermination on the Hubble rate and $\Omega_{\mathrm{X}0}$ is the 
critical fraction of a given species.}  $\omega_{\mathrm{b}0} \ll \omega_{\mathrm{M}0}$ the baryon-radiation 
contribution is suppressed as $\omega_{\mathrm{b}0}/\omega_{\mathrm{c}0}$.
After neutrino decoupling
the neutrino fraction $R_{\nu} = \rho_{\nu}/\rho_{\mathrm{R}}$ is smaller than the photon fraction 
$R_{\gamma}= \rho_{\gamma}/\rho_{\mathrm{R}}=1 - R_{\nu}$:
\begin{equation}
R_{\nu} = \frac{r}{1 + r},\qquad 
r = \frac{7}{8} N_{\nu} \biggl(\frac{4}{11}\biggr)^{4/3} \equiv 0.681 \biggl(\frac{N_{\nu}}{3}\biggr).
\label{Rnu}
\end{equation}
The third entropic fluctuation  resides then in the neutrino-photon sector, i.e.
\begin{equation}
{\mathcal S}_{\nu\gamma}= - 3 (\zeta_{\nu} - \zeta_{\gamma}) = \frac{3}{4}(\delta_{\gamma} - \delta_{\nu}).
\label{NAD3}
\end{equation}
A further solution of the truncated Einstein-Boltzmann hierarchy consists in initializing 
the lowest multipoles by assuming that the neutrino and baryon-photon dipoles are non-vanishing 
but still satisfy the momentum constraint of Eq. (\ref{MOM}). 
This solution is generically known as the neutrino-velocity mode.
Finally, in the terminology of Eqs. (\ref{NAD2}) and (\ref{NAD3}) the  adiabatic mode arises when all ${\mathcal S}_{\mathrm{ij}}$ vanishes identically, 
which implies, in the pre-equality plasma, that $\zeta_{\gamma} = \zeta_{\nu} = \zeta_{\mathrm{b}} = \zeta_{\mathrm{c}}$.

As pointed out in the past \cite{KS} (see also 
\cite{peeiso})  
the most general initial data of the Boltzmann 
hierarchy can be summarized, in the absence of a magnetized 
contribution, by a $5\times 5$ matrix \cite{tur1,chal} where the dominant amplitude might be the one associated with the adiabatic mode. 
This approach stimulated, in recent years, various attempts of including non-adiabatic contributions on the CMB initial data \cite{H4,H5} (see also \cite{H1,H2,H3}). 

The  Boltzmann hierarchy can be
 initialized by a mixture of modes: one of the modes can be adiabatic and others non-adiabatic. Predominantly adiabatic initial conditions 
 are the ones where the adiabatic amplitude is larger than the 
 various non-adiabatic amplitudes.  The strategy 
 defined in the two previous sentences summarizes the bottom-up approach to initial conditions. Combinations of one (dominant) adiabatic mode and of other (subleading) 
modes are not excluded but sometimes even help in improving the overall 
fit to the cosmological observables \cite{H3}. If, on top of the adiabatic solution, there are two or three non-adiabatic modes then 
it is hard, with the present data, to infer stringent constraints 
on the non-adiabatic components.

Why is it important  to include consistently magnetic fields
when setting initial conditions for the adiabatic and non-adiabatic modes? 
The answer on the nature of large-scale magnetization 
in the present Universe is still under active discussion \cite{review1,review2,review3}. 
The scrutiny of CMB initial conditions is a powerful window on the 
possible existence of large-scale magnetic fields  in the pre-equality 
stage.  If the origin of the large-scale magnetic fields is 
 primordial (as opposed to astrophysical) it is plausible to 
  expect the presence of magnetic fields in the primeval plasma also  before 
 the decoupling of radiation from matter.  
 CMB anisotropies
 are germane to several aspect of large-scale magnetization (see, for instance, 
 \cite{kandu,maxmagn} for two topical reviews on the subject).
 In the recent past valuable discussions of the role of pre-equality 
 magnetic fields have been conducted in different frameworks.
 For instance fully covariant approaches have been used  
to characterize the interplay between large-scale 
magnetic fields and curvature perturbations in their relativistic 
regime \cite{cov1,cov2} (see also \cite{review3}). In other studies 
the vector and the tensor modes induced by large-scale magnetic fields 
have been more specifically addressed \cite{k1,t1}. 
The approach pursued in this paper, directly linked to 
the theoretical framework of \cite{mg1,mg2,mg3},  completes and extends earlier 
works insofar as previous studies did not address specifically the 
calculation of the standard CMB observables for different 
initial conditions of the Einstein-Boltzmann hierarchy. The 
present approach wants to fill this gap and 
bring the analysis of magnetized CMB anisotropies 
to the same standard of the conventional case 
where large-scale magnetic fields are not included.
Of course, in the past, various interesting effects 
have been pointed out for specific magnetic field configurations. 
Elegant formalisms have been also explored.
While this was not a useless exercise, the quality of the present observational data 
clearly demands new and more sound theoretical calculations both 
at the analytical and numerical level. At the same time 
the improved theoretical and numerical tools should also delicately 
improve on the case where magnetic fields are absent: in the opposite 
case the comparison with the putative $\Lambda$CDM model (and its possible 
improvement) would be much more cumbersome.

The plan of the present paper is then be the following. In Section \ref{sec2} the truncated Einstein-Boltzmann hierarchy 
is discussed in the presence of large-scale magnetic fields. In Section \ref{sec3} 
the magnetized adiabatic mode is swiftly reviewed. 
Sections \ref{sec4}, \ref{sec5} and \ref{sec6} will be devoted, respectively, to the magnetized CDM-radiation mode, to the magnetized 
baryon-radiation mode and to the entropic mode of the neutrino sector. In Section \ref{sec7} 
we will discuss the case of mixtures of initial data when the magnetized adiabatic mode dominates over
the other (non-adiabatic) contributions. Concluding discussions are collected in Section \ref{sec8}. Without indulging in idle details,
the relevant analytical tools have been summarized, within the 
synchronous gauge and within our set of conventions,  in the Appendix \ref{APPA}. This choice 
makes the present script reasonably self-contained.

 \renewcommand{\theequation}{2.\arabic{equation}}
\setcounter{equation}{0}
\section{Initial data of the  Einstein-Boltzmann hierarchy}
\label{sec2}
The Einstein-Boltzmann hierarchy, appropriately truncated to the lowest multipoles, is initialized by complying with various intermediate steps:
\begin{itemize} 
\item{} the baryon-photon evolution equations have to be solved  the tight-coupling 
approximation\footnote{The tight-coupling expansion implies, to zeroth-order,  that the photon-baryon velocity are equal (see \cite{peeyu} for the pioneering work on the semi-analytical description of scalar CMB anisotropies in the tight coupling limit). This 
approximation is used, at early times, also in the  Boltzmann solvers 
to avoid numerical instabilities related to the stiffness of the problem.};
\item{} the neutrino hierarchy should be truncated to the quadrupole (or, depending 
upon the specific mode, to the octupole);
\item{} after solving the evolution equations 
for the CDM evolution, the metric fluctuations 
can be computed (for instance in the synchronous gauge).
\end{itemize}
The momentum and Hamiltonian constraints 
(stemming, respectively, from the $(0i)$ and from the $(00)$ components 
of the perturbed Einstein equations) have to be consistently 
enforced on the set of initial data.
The Einstein-Boltzmann hierarchy must be initialized
deep in the radiation-dominated stage of expansion and for typical wavelengths larger than the Hubble radius at the corresponding epoch. The two mentioned physical limits define, as we shall see in a moment, the relevant expansion parameters of the problem.
The solution of the background Friedmann-Lema\^itre 
equations (i.e. Eq. (\ref{FL})) across the radiation-matter transition (and for a spatially flat Universe) reads
\begin{equation}
\alpha = \frac{a}{a_{\mathrm{eq}}}= x^2 + 2 x, \qquad x = \frac{\tau}{\tau_{1}}, \qquad \tau_{1} = \frac{2}{H_{0}} \sqrt{\frac{a_{\mathrm{eq}}}{\Omega_{\mathrm{M}  0}}}  \simeq 288 \,\, \biggl(\frac{ \omega_{\mathrm{M}0}}{0.134}\biggr)^{-1}\, \mathrm{Mpc},
\label{alpha}
\end{equation}
where $a_{\mathrm{eq}}$ is the scale factor at the equality, i.e. the 
moment when non-relativistic matter and radiation contribute equally to the 
total energy density of the plasma.
For  $\alpha \ll 1$ (i.e. $a \ll a_{\mathrm{eq}}$) the plasma is dominated by radiation, and, according to Eq. (\ref{alpha}),  $\alpha \simeq 2 x  + {\mathcal O}(x^2)= 2 (\tau/\tau_{1})$. Furthermore, as it 
can be easily appreciated, $\alpha = \rho_{\mathrm{M}}/ \rho_{\mathrm{R}}$. Slightly different 
conventions can be adopted \footnote{In the literature there is sometimes the habit 
of setting $\tau_{1} = 1/2 $ and $a_{\mathrm{eq}}= 1/4$. This choice would
imply that $a(\tau) = \tau + \tau^2$. In the latter case the asymptotic 
solution during radiation will go as $a(\tau) \simeq \tau$. Thus the multiplicative factors of the various multipoles 
will change in comparison with the conventions of the present paper. In the numerical code, the evolution of the scale factor is computed numerically as a function of $\tau$ measured in comoving units 
of Mpc (while $k$, the wavenumber, will be measured in units of $\mathrm{Mpc}^{-1}$). }.

Defining as $\tau_{\mathrm{i}}$ the initial integration time (which will be after neutrino decoupling but before 
matter-radiation equality) it will be required that $k \tau_{\mathrm{i}} < 1$ for all the modes involved in the calculations. 
The smallness of $k \tau_{\mathrm{i}}$ measures the excess 
of the given wavelength with respect to the Hubble radius at the time 
$\tau_{\mathrm{i}}$ (typically selected well before $\tau_{\mathrm{eq}}$).
Summarizing, the double expansion employed in setting initial conditions
of the truncated Einstein-Boltzmann hierarchy can be formally written as
\begin{equation}
\alpha \ll 1, \qquad \frac{k}{ a H} = \frac{k}{ {\mathcal H}} = \frac{\kappa \,\alpha}{2\,\sqrt{\alpha + 1}} \simeq 
k \tau \ll 1, \qquad {\mathcal H} = \frac{2}{\tau_{1}} \frac{\sqrt{\alpha + 1}}{\alpha}.
\label{expans1}
\end{equation}
where $\kappa = k \tau_{1}$ measures how large the wavelength was, in Hubble units, around equality (note, indeed, that $\tau_{\mathrm{eq}} = (\sqrt{2} -1) \tau_{1} \simeq \tau_{1}/2$).
To derive the expression of the various adiabatic and non-adiabatic modes it is practical to use, as a guiding 
principle, the evolution of curvature perturbations in the limit $ k/{\mathcal H} \simeq k\tau \to 0$. 
The obtained expression will be used (to leading order in the $\alpha$-expansion) for solving the truncated Einstein-Boltzmann hierarchy to a given order in $k\tau$.
The two mentioned steps can be iterated at wish to obtain the initial conditions with the wanted accuracy either in $k\tau$ or in $\alpha$.

Before plunging into the discussion it is appropriate to recall 
\cite{mg2,giokunz2} that large-scale magnetic fields 
affect directly the Einstein equations since they contribute to the energy density, to the pressure and to the anisotropic stress:
\begin{equation}
\delta_{\mathrm{s}} \rho_{\mathrm{B}}(\vec{x},\tau) = \frac{B^{2}(\vec{x},\tau)}{8\pi a^4(\tau)},\qquad \delta_{\mathrm{s}} p_{\mathrm{B}}(\vec{x},\tau) = \frac{\delta_{\mathrm{s}} \rho_{\mathrm{B}}(\vec{x},\tau)}{3},
\qquad \tilde{\Pi}_{i}^{j}(\vec{x},\tau) = 
\frac{1}{4\pi a^4(\tau)} \biggl( B_{i}B^{j} - \frac{B^2}{3} \delta_{i}^{j}\biggr),
\label{MAGN1}
\end{equation}
where $B^2 = B_{i}B^{i}$. For notational convenience, and 
in analogy with what customarily done to treat 
the neutrino anisotropic stress \cite{cosmics2},  we will
also define 
\begin{equation}
\partial_{j}\partial^{i} \tilde{\Pi}_{i}^{j} = (p_{\gamma} + \rho_{\gamma}) \nabla^2 \sigma_{\mathrm{B}},\qquad \Omega_{\mathrm{B}}(\vec{x},\tau) 
= \frac{\delta_{\mathrm{s}} \rho_{\mathrm{B}}(\vec{x},\tau)}{\rho_{\gamma}(\tau)},
\label{MAGN2}
\end{equation}
where $\sigma_{\mathrm{B}}$ and $\Omega_{\mathrm{B}}$ are both dimensionless.
The magnetic component also enter directly the Boltzmann 
hierarchy. In the truncated version, which is used to set initial conditions 
the magnetohydrodynamical (MHD) approximation,  the Lorentz force 
affects the baryon-lepton-photon evolution. In this respect it is useful 
to bear in mind that 
\begin{equation}
\frac{\vec{\nabla}\cdot({\vec{J}\times \vec{B})}}{a^4 \rho_{\mathrm{b}}} = 
\frac{1}{R_{\mathrm{b}}} \biggl[ \nabla^2 \sigma_{\mathrm{B}} - \frac{1}{4} 
\nabla^2 \Omega_{\mathrm{B}}\biggr],
\label{MAGN3}
\end{equation}
where $\vec{J} \simeq \vec{\nabla}\times \vec{B}/(4\pi)$ is the Ohmic 
current  and $R_{\mathrm{b}}$ is the baryon-to-photon ratio which can be 
expressed as 
\begin{equation}
R_{\mathrm{b}} = \frac{3}{4} \frac{\rho_{\mathrm{b}}}{\rho_{\gamma}} =
\frac{3}{4 R_{\gamma}} \frac{\Omega_{\mathrm{M}}}{\Omega_{\mathrm{R}}} \frac{\omega_{\mathrm{b0}}}{\omega_{\mathrm{M}0}} = \frac{3\alpha}{4 R_{\gamma}} \frac{\omega_{\mathrm{b0}}}{\omega_{\mathrm{M}0}},\qquad \Omega_{\mathrm{R}} = \frac{1}{1+ \alpha},\qquad \Omega_{\mathrm{M}} = \frac{\alpha}{1 + \alpha},
\label{MAGN4}
\end{equation}
where, by definition, $R_{\gamma} = \rho_{\gamma}/\rho_{\mathrm{R}}$.
The relevance of $R_{\mathrm{b}}$ depends on the specific solution. 
As it can be also appreciated from Eq. (\ref{thetab}), the contribution 
of $R_{\mathrm{b}}$ is always suppressed, before equality, since 
it is proportional both to $\alpha \ll 1$ and to $\omega_{\mathrm{b}0}/\omega_{\mathrm{M}0}$. However, to get accurate initial conditions 
it is often important to take into account the contribution of $R_{\mathrm{b}}$ to the asymptotic form of the evolution
equation of the baryon-photon dipole.
To leading order in $k/{\mathcal H}$ the evolution of the curvature 
perturbations\footnote{The evolution equation of the curvature perturbations is swiftly derived, for 
completeness, in Eq. (\ref{der3}) of Appendix \ref{APPA}.} reads \cite{mg1,mg2}:
\begin{equation}
{\mathcal R}' = - {\mathcal H} \frac{\delta_{\mathrm{s}} p_{\mathrm{nad}}}{p_{\mathrm{t}} + \rho_{\mathrm{t}}} + 
\frac{{\mathcal H}}{p_{\mathrm{t}} + \rho_{\mathrm{t}}} \biggl( c_{\mathrm{st}}^2 - \frac{1}{3} \biggr) 
\delta_{\mathrm{s}} \rho_{\mathrm{B}} + {\mathcal O}\biggl(\frac{k^2}{{\mathcal H}^2}\biggr).
\label{Rev}
\end{equation}
The remaining terms appearing in Eq. (\ref{der3}) will be, as 
specified in Eq. (\ref{Rev}), at most, ${\mathcal O}(k^2\tau^2)$ except for the peculiar case of the neutrino-velocity mode where they will be, at most, ${\mathcal O}(k\tau)$.  Once the solution of ${\mathcal R}$ is known, the obtained 
result can be used in the following equation:
\begin{equation}
\frac{{\mathcal H}}{{\mathcal H}^2 - {\mathcal H}'} \xi' + \xi = {\mathcal R}.
\label{xiev}
\end{equation}
Equation (\ref{xiev}), read from right to left, is just the definition of ${\mathcal R}$ in terms of the 
synchronous degrees of freedom.  However, if ${\mathcal R}$ is determined from Eq. (\ref{Rev}), Eq. (\ref{xiev}) 
allows to determine $\xi$.
The solution of (\ref{xiev}) can finally be inserted into the Hamiltonian constraint, i.e. Eq. (\ref{HAM}):
\begin{equation}
2 k^2 \xi - {\mathcal H} h' = 8\pi G a^2 [ \delta_{\mathrm{s}} \rho_{\mathrm{t}} + \delta_{\mathrm{s}}\rho_{\mathrm{B}}],
\label{hev}
\end{equation}
to determine $h(k,\alpha)$. Since both Eqs. (\ref{xiev}) and (\ref{hev}) 
are exact, the solution of Eq. (\ref{hev}) even allows for the determination
of $h(k,\alpha)$  to first order in $k^2 \tau^2$.
Across the radiation-matter transition,  the total barotropic index and the 
total sound speed can be written, respectively, as
\begin{equation}
w_{\mathrm{t}} = \frac{p_{\mathrm{t}}}{\rho_{\mathrm{t}}} = \frac{1}{3(\alpha + 1)}, \qquad 
c_{\mathrm{st}}^2 = \frac{\partial p_{\mathrm{t}}}{\partial \rho_{\mathrm{t}}} = \frac{4 }{3 ( 3 \alpha + 4)}.
\label{ss}
\end{equation}
Therefore, Eqs. (\ref{Rev}) and (\ref{xiev}) become, in explicit terms,
\begin{eqnarray}
&& \frac{\partial {\mathcal R}}{\partial \alpha} = - \frac{4{\mathcal D}_{*}(k)}{(3 \alpha + 4)^2} ,
\label{Rev2}\\
&& \frac{\partial \xi}{\partial \alpha} + \frac{3 \alpha + 4}{2 \alpha (\alpha + 1)} \xi = \frac{3 \alpha + 4}{2 \alpha (\alpha + 1)} {\mathcal R},
\label{xiev2}
\end{eqnarray}
where, for short,  the quantity 
\begin{equation}
{\mathcal D}_{*}(k) = \biggl[ \frac{\omega_{\mathrm{c}0}}{\omega_{\mathrm{M}0}} {\mathcal S}_{*}(k) + 
\frac{3}{4} R_{\gamma} \Omega_{\mathrm{B}}(k)  \biggr],
\label{Ddef}
\end{equation}
has been introduced.
In Eq. (\ref{Rev2}) only one CDM-radiation 
mode has been included as an illustration of the procedure
and, in this case,
\begin{equation}
 \delta_{\mathrm{s}} p_{\mathrm{nad}}= \rho_{\mathrm{c}}
c_{\mathrm{st}}^2 {\mathcal S}_{*}(k),\qquad {\mathcal S}_{*}(k) =  \frac{3}{4} \delta_{\mathrm{r}}(k,\tau) - \delta_{\mathrm{c}}(k,\tau). 
\label{Sdef}
\end{equation}
 Equations (\ref{Rev2}) and (\ref{xiev2}) can then be integrated once 
 with respect to $\alpha$; the result is:
\begin{eqnarray}
&&{\mathcal R}(k,\alpha) = {\mathcal R}_{*}(k) - \frac{\alpha {\mathcal D}_{*}(k)}{3\alpha +4} ,
\label{Rsol1}\\
&& \xi(k,\alpha) = {\mathcal R}_{*}(k) - \frac{{\mathcal D}_{*}(k)}{3 \alpha^2} [ \alpha ( \alpha - 4) + 8 (\sqrt{\alpha +1} -1)],
\label{xisol1}
\end{eqnarray}

In the long wavelength limit defined by Eq. (\ref{expans1}), the density contrasts following from 
Eqs. (\ref{deltanu})--(\ref{deltac}) 
\begin{eqnarray}
&&\delta_{\nu}(k,\alpha) \simeq - R_{\gamma} \Omega_{\mathrm{B}}(k) + \frac{2}{3} h(k,\alpha),
\label{deltanuex1}\\
&& \delta_{\gamma}(k,\alpha) \simeq - R_{\gamma} \Omega_{\mathrm{B}}(k) + \frac{2}{3} h(k,\alpha),
\label{deltagammaex1}\\
&& \delta_{\mathrm{b}}(k,\alpha) \simeq - \frac{3}{4} R_{\gamma} \Omega_{\mathrm{B}}(k) + \frac{h(k,\alpha)}{2},
\label{deltabex1}\\
&& \delta_{\mathrm{c}}(k,\alpha) \simeq - {\mathcal S}_{*}(k)
- \frac{3}{4} R_{\gamma} \Omega_{\mathrm{B}}(k) + \frac{h(k,\alpha)}{2}.
\label{deltacex1}
\end{eqnarray}
If ${\mathcal S}_{*}(k)=0$ in Eq. (\ref{deltacex1}), Eqs. (\ref{deltanuex1})--(\ref{deltacex1}) 
will describe the adiabatic mode. If ${\mathcal S}_{*}(k)\neq 0$ in Eq. (\ref{deltacex1}), Eqs. (\ref{deltanuex1})--(\ref{deltacex1}) apply in the case where the non-adiabatic contribution is present either 
in combination with the adiabatic mode or in the absence of the adiabatic component. 

By inserting Eq. (\ref{xisol1})  into Eq. (\ref{hev}) (written in the form given in Eq. (\ref{HAM1}))  a decoupled equation for $h(k,\alpha)$ 
can be obtained with the combined use of Eqs. (\ref{deltanuex1})--(\ref{deltacex1}):
\begin{equation}
\frac{\partial h}{\partial \alpha} + \frac{3 \alpha + 4}{2 \alpha (\alpha + 1)} h = \frac{3 {\mathcal D}_{*}(k)}{\alpha+1} + \frac{2 k^2 \xi}{{\mathcal H}^2 \alpha}.
\label{heqex2}
\end{equation}
Equation (\ref{heqex2}) can then be solved with elementary methods and the result is:
\begin{equation}
h(k,\alpha) = {\mathcal D}_{*}(k) {\mathcal F}_{1}(\alpha) + \kappa^2 {\mathcal R}_{*}(k) {\mathcal F}_{2}(\alpha) - \kappa^2 {\mathcal D}_{*}(k) {\mathcal F}_{3}(\alpha),
\label{hsol1}
\end{equation}
where
\begin{eqnarray}
{\mathcal F}_{1}(\alpha) &=& 2\frac{[ \alpha (\alpha - 4) + 8 (\sqrt{\alpha +1} -1)]}{\alpha^2},
\label{F1}\\
{\mathcal F}_{2}(\alpha) &=&\frac{[ \alpha^3 - 2 \alpha^2 + 8\alpha + 16 ( 1 - \sqrt{\alpha + 1})]}{{5 \alpha^2} },
\label{F2}\\
{\mathcal F}_{3}(\alpha) &=& \frac{\{ 32 (\sqrt{\alpha + 1} -1) + 
\alpha [ \alpha ( 3 \alpha - 26) + 60 \sqrt{\alpha + 1} - 16] - 60 \sqrt{\alpha + 1} \ln{(\alpha + 1)}\}}{{45 \alpha^2} }.
\label{F3}
\end{eqnarray}
\begin{figure}[!ht]
\centering
\includegraphics[height=6.3cm]{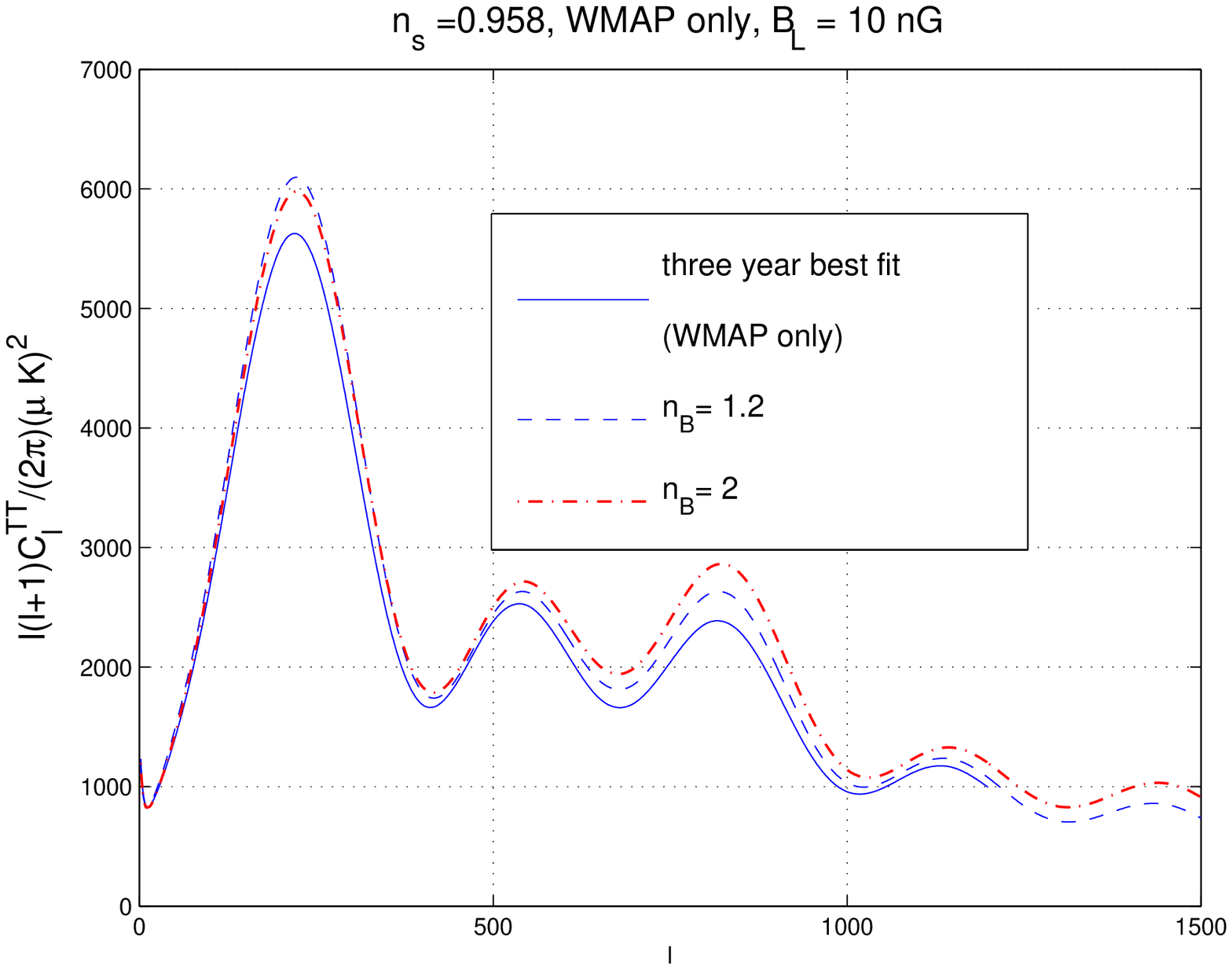}
\includegraphics[height=6.3cm]{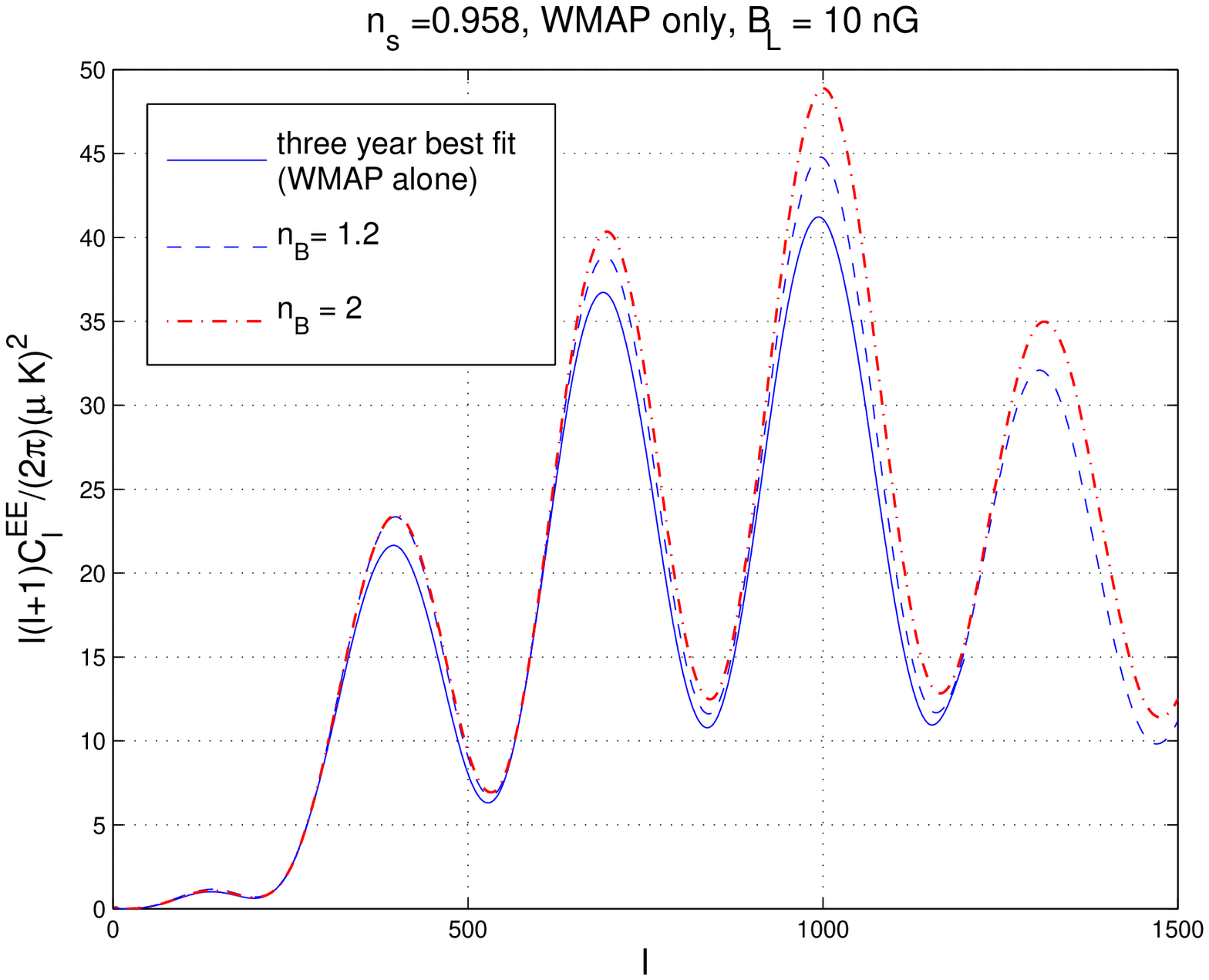}
\includegraphics[height=6.3cm]{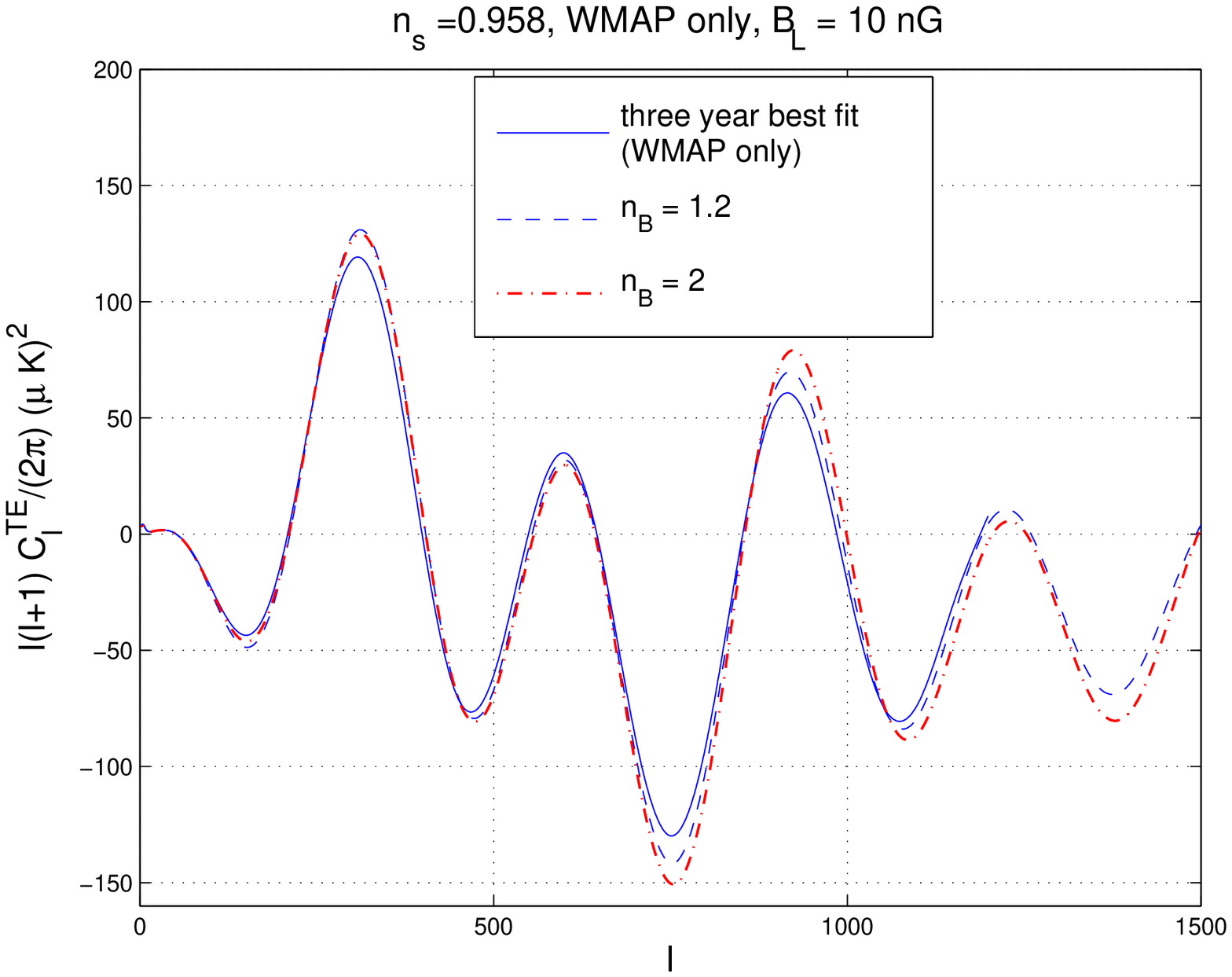}
\includegraphics[height=6.3cm]{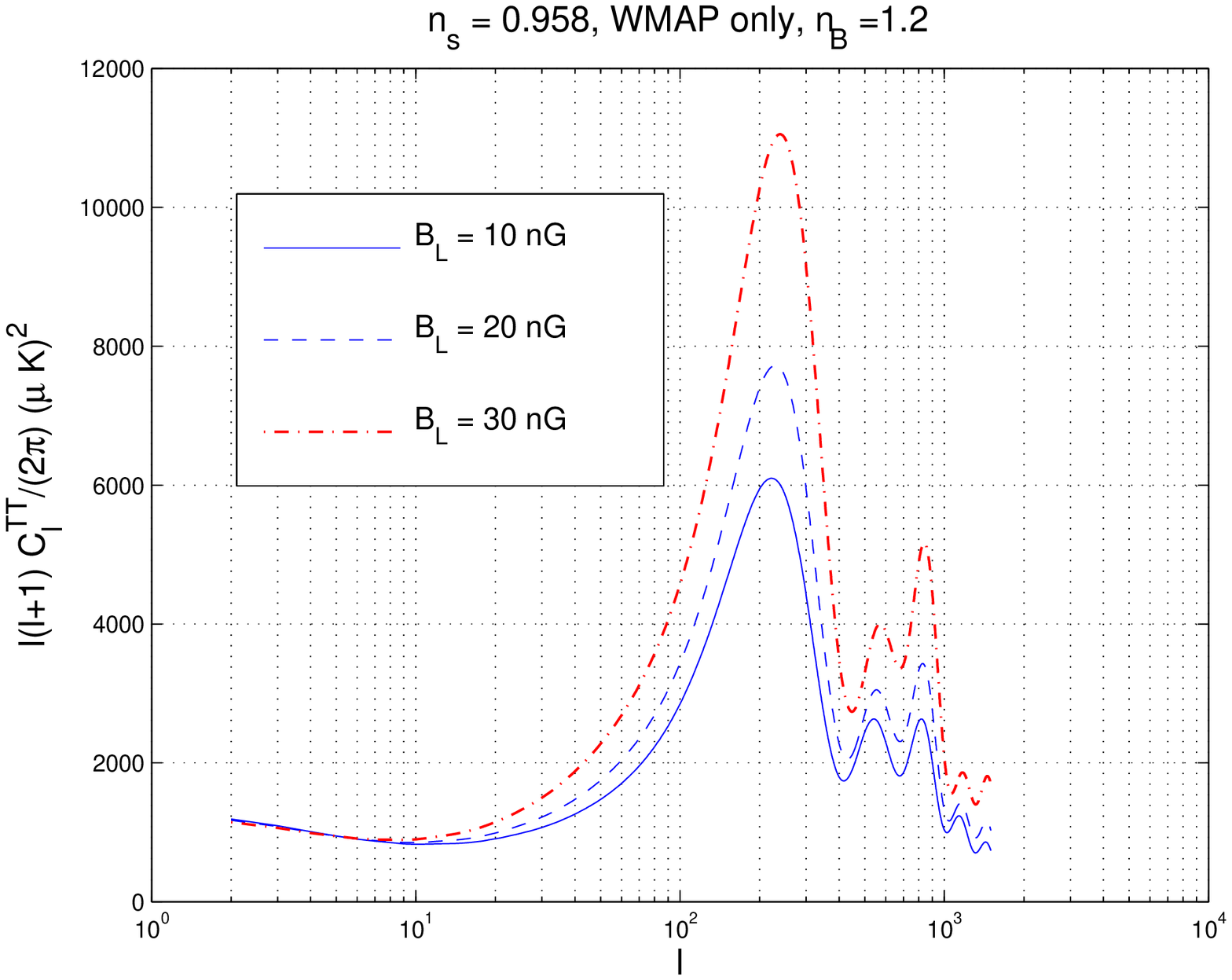}
\caption[a]{The magnetized adiabatic mode for spatially flat models is illustrated. The $\Lambda$CDM 
parameters are selected to match with the ones 
of the WMAP three year best fit to the WMAP alone (i.e. 
$h_{0}= 0.732$, $\omega_{\mathrm{b}0} =0.02229$, 
$\omega_{\mathrm{c}0} =0.1054$ and ${\mathcal A}_{{\mathcal R}}=2.35 \times 10^{-9}$). The combination of the WMAP data 
with the other sets of cosmological data (or with the other  CMB experiments) leads to best fit parameters which are 
largely compatible with the ones used here only for sake 
of illustration.}
\label{Figure1}      
\end{figure}
In the absence of magnetized contribution  ${\mathcal R}(k,\alpha)$ is
constant, to leading order in $\alpha$ and $k\tau$, provided ${\mathcal S}_{*}(k)=0$. This is also 
true for $\xi(k,\alpha)$. The situation for $h(k,\alpha)$ is a bit different. 
Indeed, the synchronous coordinate system implies that the gauge freedom is not completely 
fixed. Thus the possible presence of a constant mode in $h(k,\alpha)$ is the reflection 
of a gauge artifact which can be removed by a further gauge transformation
which will keep the transformed fluctuation always within the synchronous coordinate system \cite{cosmics1,syn1,bardeen}.  In our approach 
the gauge mode does not show up since we integrated directly the 
evolution equation of curvature perturbations (i.e. Eq. (\ref{Rev}))
which is gauge-invariant and hence free of possible gauge modes.
By selecting a given order in $\alpha$, the asymptotic solution can be fed back into all the other equations of the truncated Einstein-Boltzmann hierarchy. The system can then be solved to a given order in $k\tau$. There is no general way of achieving 
the second step. It will thus be mandatory to solve the system separately 
for the different modes.

\renewcommand{\theequation}{3.\arabic{equation}}
\setcounter{equation}{0}
\section{Magnetized adiabatic mode}
\label{sec3}
The long wavelength solution derived in the previous section 
implies that, to leading order, 
\begin{equation}
\xi(k,\tau) \simeq {\mathcal R}_{*}(k) + {\mathcal O}(k^2\tau^2), \qquad  \zeta_{\gamma}(k,\tau) \simeq \zeta_{\nu}(k,\tau) \simeq \zeta_{\mathrm{c}}(k,\tau)
\simeq \zeta_{\mathrm{b}}(k,\tau).
\label{MAD1}
\end{equation}
Thus, to leading order in $\alpha$ and in $k\tau$ 
the ${\mathcal S}_{\mathrm{ij}}$ introduced in  Eqs. (\ref{DPNAD1}) and (\ref{DPNAD2}) are all vanishing.
 The result of the integration 
of the truncated Einstein-Boltzmann hierarchy of Appendix 
\ref{APPA} can then be written, in Fourier space, as:
\begin{eqnarray}
 \xi(k,\tau) &=& {\mathcal R}_{*}(k) + \biggl\{- \frac{4 R_{\nu} + 5}{12( 4 R_{\nu} + 15)} {\mathcal R}_{*}(k) +\frac{R_{\gamma} [ 4 \sigma_{\mathrm{B}}(k) - R_{\nu} \Omega_{\mathrm{B}}(k)]}{ 6 ( 4 R_{\nu} + 15)}  \biggr\} k^2 \tau^2,
\label{AD1}\\
h(k,\tau) &=& \frac{{\mathcal R}_{*}(k)}{2} k^2 \tau^2 - \frac{1}{36} \biggl\{- \frac{8 R_{\nu}^2 - 14 R_{\nu} - 75}{2(2 R_{\nu} + 25)(4 R_{\nu} + 15)} {\mathcal R}_{*}(k)
\nonumber\\
&+& \frac{R_{\gamma} ( 15 - 20 R_{\nu})}{10( 4 R_{\nu} + 15) ( 2 R_{\nu} + 25)} [R_{\nu}\Omega_{\mathrm{B}}(k) - 4 \sigma_{\mathrm{B}}(k)]\biggr\} k^4 \tau^4,
\label{AD2}\\
\delta_{\gamma}(k,\tau) &=& - R_{\gamma} \Omega_{\mathrm{B}}(k) + \frac{2}{3} \biggl[ 
\frac{{\mathcal R}_{*}(k)}{2} + \sigma_{\mathrm{B}}(k) - \frac{R_{\nu}}{4} \Omega_{\mathrm{B}}(k)\biggr] k^2 \tau^2,
\label{AD33}\\
\delta_{\nu}(k,\tau) &=& - R_{\gamma} \Omega_{\mathrm{B}}(k) + \frac{2}{3} \biggl[ \frac{{\mathcal R}_{*}(k)}{2}  - \frac{R_{\gamma}}{ R_{\nu}} \sigma_{\mathrm{B}}(k) - \frac{R_{\gamma}}{4} \Omega_{\mathrm{B}}(k)\biggr] k^2 \tau^2,
\label{AD4}\\
\delta_{\mathrm{c}}(k,\tau) &=& - \frac{3}{4}R_{\gamma} \Omega_{\mathrm{B}}(k) + \frac{{\mathcal R}_{*}(k)}{4} k^2 \tau^2,
\label{AD5}\\
\delta_{\mathrm{b}}(k,\tau) &=& - \frac{3}{4}R_{\gamma} \Omega_{\mathrm{B}}(k) + \frac{1}{2} \biggl[ \frac{{\mathcal R}_{*}(k)}{2}  + \sigma_{\mathrm{B}}(k)- \frac{R_{\nu}}{4} \Omega_{\mathrm{B}}(k) \biggr] k^2\tau^2,
\label{AD6}\\
\theta_{\gamma\mathrm{b}}(k,\tau) &=& \biggl[ \frac{R_{\nu}}{4} \Omega_{\mathrm{B}}(k) - \sigma_{\mathrm{B}}\biggr] 
k^2 \tau -\frac{1}{36} \biggl[ -{\mathcal R}_{*}(k) + \frac{R_{\nu} \Omega_{\mathrm{B}}(k) - 4 \sigma_{\mathrm{B}}(k)}{2}\biggr] k^4\tau^3,
\label{AD7}\\
\theta_{\nu}(k,\tau) &=& \biggl[ \frac{R_{\gamma}}{R_{\nu}} \sigma_{\mathrm{B}}(k) - \frac{R_{\gamma}}{4} \Omega_{\mathrm{B}}(k)\biggr] k^2 \tau
- \frac{1}{36}\biggl\{-\frac{( 4 R_{\nu} + 23)}{4 R_{\nu} + 15} {\mathcal R}_{*}(k) 
\nonumber\\
&+& \frac{R_{\gamma}( 4 R_{\nu} + 27)}{2 R_{\nu} ( 4 R_{\nu} + 15)}[ 4 \sigma_{\mathrm{B}}(k) - R_{\nu} \Omega_{\mathrm{B}}(k)]\biggr\} k^4 \tau^3,
\label{AD8}\\
 \theta_{\mathrm{c}}(k,\tau) &=& 0,
\label{AD9}\\
\sigma_{\nu}(k,\tau) &=& - \frac{R_{\gamma}}{R_{\nu}} \sigma_{\mathrm{B}}(k) + \biggl\{ -\frac{2 {\mathcal R}_{*}(k)}{3( 4 R_{\nu} + 15)} + \frac{R_{\gamma}[ 4 \sigma_{\mathrm{B}}(k) - R_{\nu} \Omega_{\mathrm{B}}(k)]}{ 2 R_{\nu}(4 R_{\nu} + 15)}\biggr\} k^2 \tau^2.
\label{AD10}
\end{eqnarray}
The adiabatic mode is often assigned 
in terms of a putative $C(k)$ which is related to ${\mathcal R}_{*}(k)$
as ${\mathcal R}_{*}(k) = - 2 C(k)$. To leading order in $k\tau$ the initial conditions are adiabatic, however, 
higher orders in $k\tau$  introduce
a perturbative mismatch between the relevant density contrasts.
Higher orders in $\alpha$ (at a given order in $k\tau$) can be easily included following the expressions of the previous section.
Since the Hamiltonian constraint has to be satisfied at the onset of the integration, the contribution of the magnetic field in the initial data 
acquires a definite sign. The relative correlation between the magnetic field and the adiabatic mode is fixed.
In Fig. \ref{Figure1} the salient features of the adiabatic mode are 
illustrated and have been more thoroughly discussed in \cite{giokunz2}.
For sake of comparison with the other sets of initial conditions, the adiabatic power spectra are assigned according to the standard 
convention, i.e. $ {\mathcal P}_{{\mathcal R}}(k) = {\mathcal A}_{\mathcal R}(k/k_{\mathrm{p}})^{n_{\mathrm{s}} -1}$,
where $k_{\mathrm{p}}=0.002\,\,\mathrm{Mpc}^{-1}$ is the pivot scale, i.e. the scale at which 
${\mathcal P}_{{\mathcal R}}(k_{\mathrm{p}}) = {\mathcal A}_{{\mathcal R}}$. The three year best fit to the WMAP data alone implies 
\cite{WMAP1,WMAP2} 
${\mathcal A}_{{\mathcal R}} =2.35 \times 10^{-9}$. According to the legends of each plot, in Fig. \ref{Figure1} the full line denotes the corresponding angular power spectra in the absence of magnetic contribution 
and for WMAP three year best fit. 

The magnetic fields are included in the simplest realization 
of what has been called in \cite{giokunz1} m$\Lambda$CDM model, i.e. 
the magnetized $\Lambda$CDM model. In this case the 
power spectra of $\Omega_{\mathrm{B}}$ and $\sigma_{\mathrm{B}}$ 
are assigned, respectively, as
${\mathcal P}_{\Omega}(k) = \overline{\Omega}_{\mathrm{BL}}^2 {\mathcal F}(n_{\mathrm{B}})(k/k_{\mathrm{L}})^{2(n_{\mathrm{B}} -1)}$ 
and as ${\mathcal P}_{\sigma}(k) = \overline{\Omega}_{\mathrm{BL}}^2 {\mathcal G}(n_{\mathrm{B}})(k/k_{\mathrm{L}})^{2(n_{\mathrm{B}} -1)}$,
where $\overline{\Omega}_{\mathrm{BL}} = B_{\mathrm{L}}^2/(8\pi
a^4\rho_{\gamma})$. The functions ${\mathcal F}(n_{\mathrm{B}})$ and 
${\mathcal G}(n_{\mathrm{B}})$ can be determined 
from the appropriate convolutions \cite{giokunz2} once the 
two-point function of the magnetic fields is known 
in Fourier space, i.e. 
\begin{equation}
\langle B_{i}(\vec{k}) B_{j}(\vec{p})\rangle = \frac{2\pi^2}{k^3} P_{ij}(k) {\mathcal P}_{\mathrm{B}}(k) \delta^{(3)}(\vec{k} + \vec{p}), \qquad P_{ij}(k) = \biggl(\delta_{ij} - \frac{k_{i} k_{j}}{k^2} \biggr).
\end{equation}
The quantity $B_{\mathrm{L}}$ is the amplitude of the
magnetic field regularized over a comoving scale $L \simeq k_{\mathrm{L}}^{-1}$. In this context $\overline{\Omega}_{\mathrm{BL}}$ measures
(up to Euler Gamma functions \cite{giokunz2}) the amplitude 
of the magnetic power spectrum ${\mathcal P}_{\mathrm{B}}(k)$ 
at the magnetic pivot scale $k_{\mathrm{L}}$.
From Fig. \ref{Figure1}, the TT spectra exhibit a distortion of the second peak which is correlated with the increase of the first and third 
acoustic peaks. As the values of the magnetic spectral index (i.e. $n_{\mathrm{B}}$) increases the distortion becomes more pronounced. As the values of the magnetic field augments beyond $20$ nG the second peak 
practically disappears and it is replaced by a sort of hump.
This result is also expected on analytical ground \cite{mg1,mg2,mg3}.
In Fig. \ref{Figure1} excessive values of the magnetic fields have been
adopted just to emphasize more visually 
the modifications of the shape which are typically related to this mode. This 
aspect can be understood  by comparing, in Fig. \ref{Figure1} the dashed (and dot-dashed) lines 
with the full lines (representing the three year best fit to the WMAP alone).
In the bottom-right plot the different curves denote three different values 
of the magnetic fields for fixed magnetic spectral index (i.e. $n_{\mathrm{B}} = 1.2$) and fixed 
adiabatic spectral index (i.e. $n_{\mathrm{s}} =0.958$).

The magnetic pivot scale 
$k_{\mathrm{L}}$ is taken (in Fig. \ref{Figure1} and in all the 
other plots of this paper) to be $1\,\, \mathrm{Mpc}^{-1}$. 
The EE and TE correlations are also modified by the presence of 
large-scale magnetic fields. The physical rationale for this occurrence can be understood already to lowest order in the tight-coupling approximation: since the 
dipole is affected both by $\sigma_{\mathrm{B}}$ and $\Omega_{\mathrm{B}}$ the TE and EE angular power spectra are modified since 
the polarization (to first-order in the tight-coupling expansion) is proportional to the zeroth-order dipole which depends, in the MHD
treatment, on the Lorentz force and, ultimately, on the Ohmic currrent. This 
aspect has been already pointed out, within a much less accurate 
semi-analytical approach, in \cite{mg1}.
\renewcommand{\theequation}{4.\arabic{equation}}
\setcounter{equation}{0}
\section{Magnetized CDM-radiation mode}
\label{sec4}
\begin{figure}[!ht]
\centering
\includegraphics[height=6.5cm]{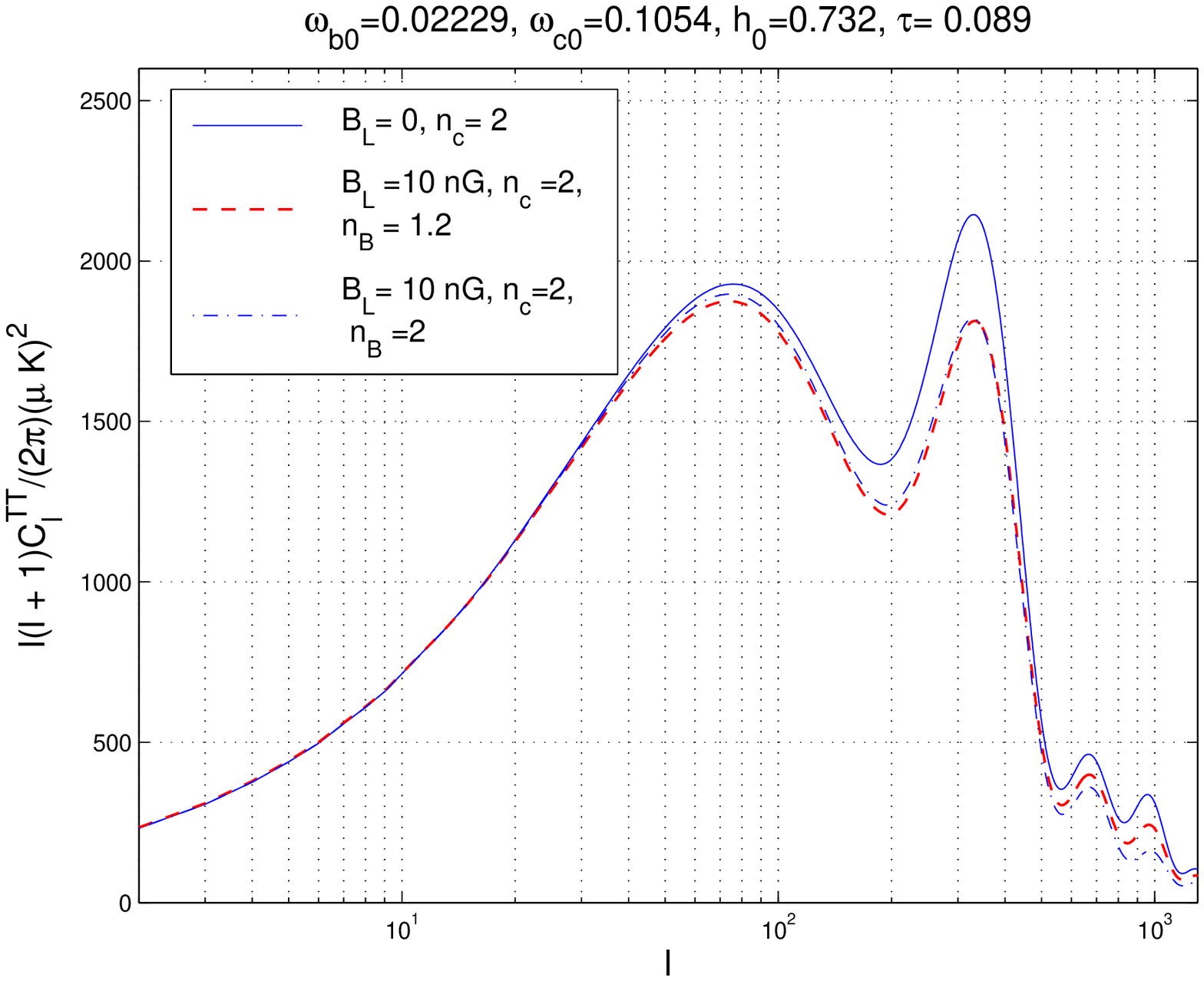}
\includegraphics[height=6.5cm]{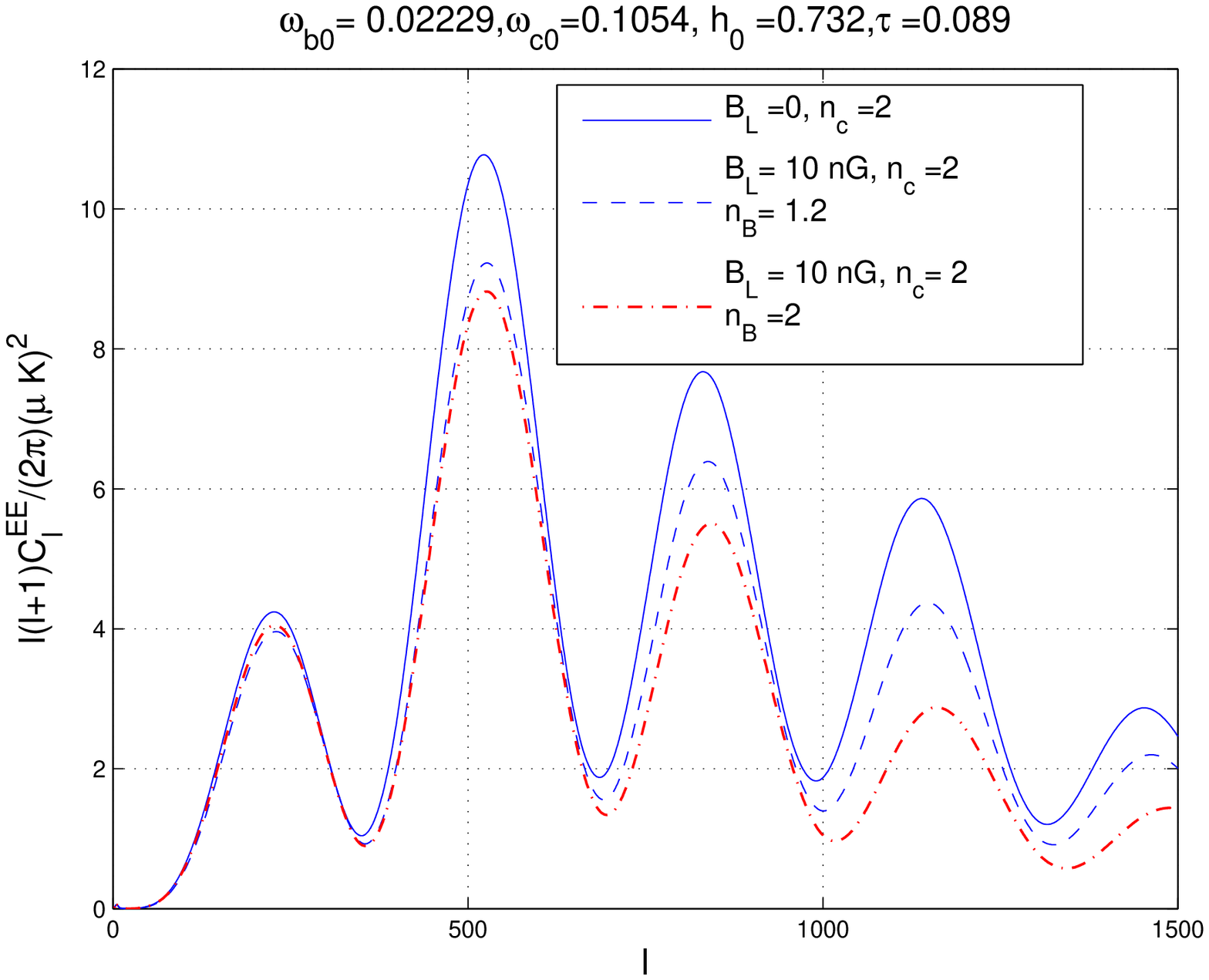}
\includegraphics[height=6.5cm]{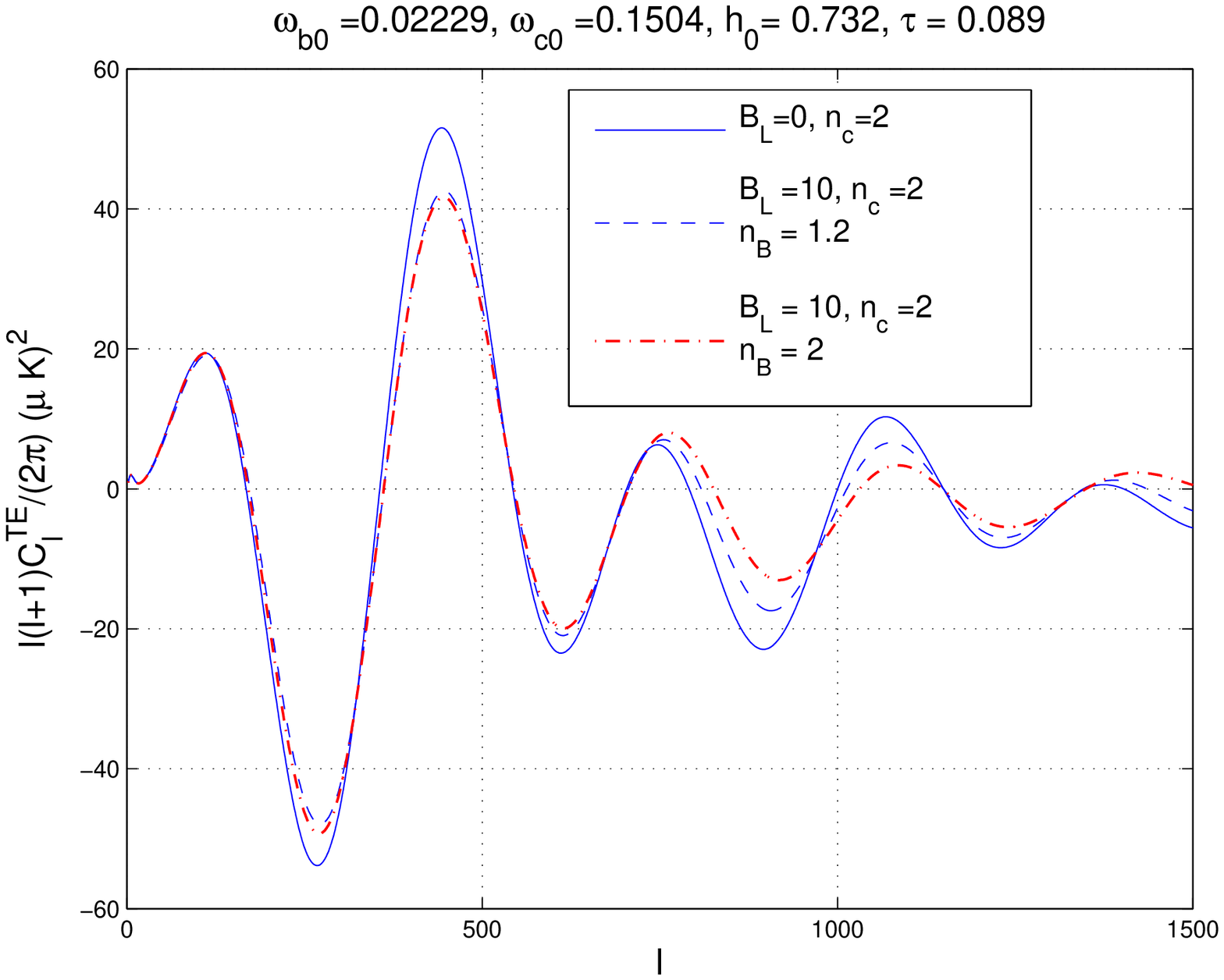}
\includegraphics[height=6.5cm]{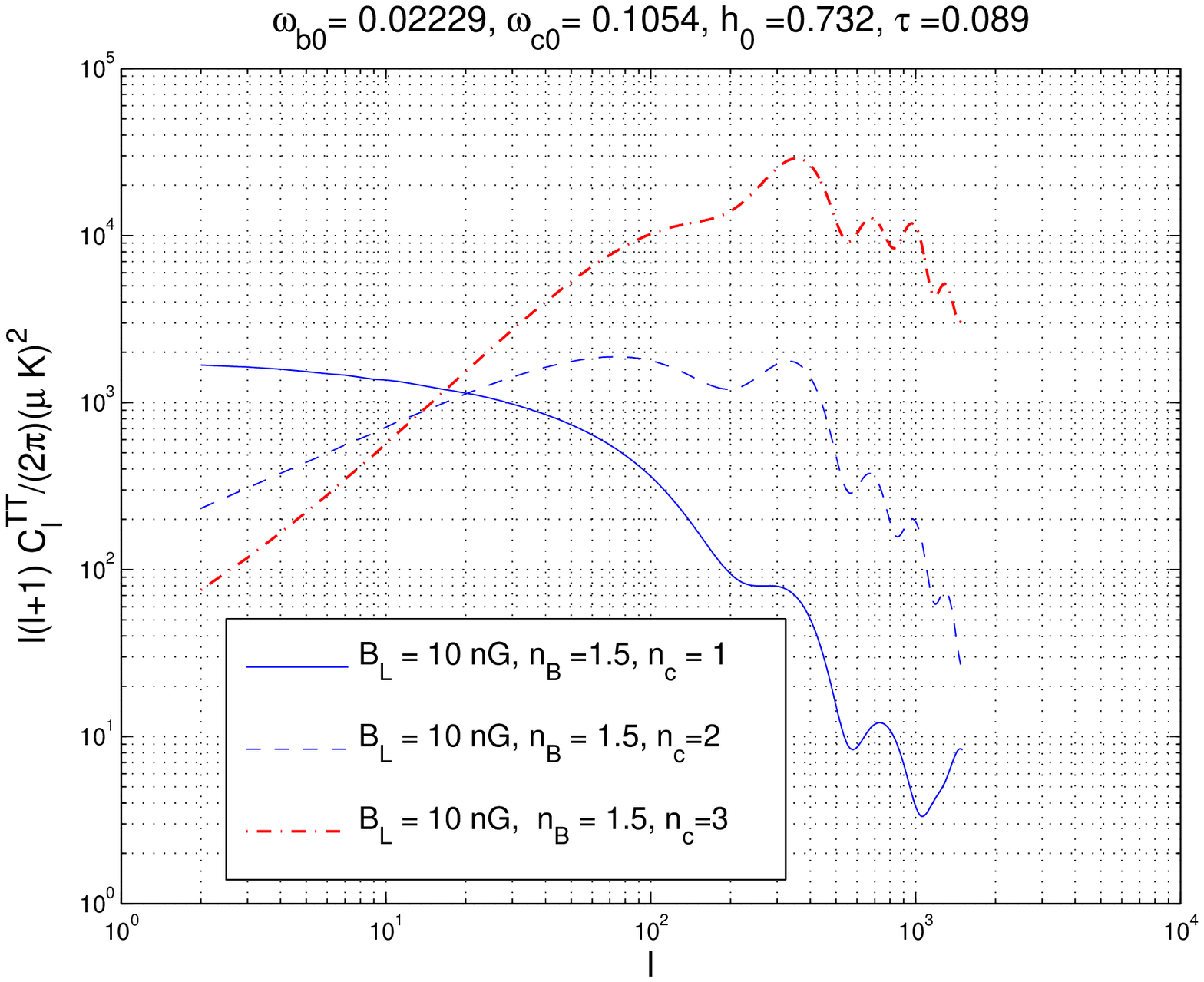}
\caption[a]{The magnetized CDM-radiation mode is illustrated for spatially flat models. The cosmological parameters, for sake of comparison, are fixed to the same values of Fig. \ref{Figure1}. }
\label{Figure2}      
\end{figure}

In the case of the magnetized CDM-radiation mode the entropy fluctuation arises in the CDM-radiation system.
Following the notations of Eq. (\ref{NAD1}) we will denote by ${\mathcal S}_{*}(k)$ the common 
amplitude, in Fourier space, of the two relevant entropy fluctuations, i.e. 
\begin{equation}
{\mathcal S}_{*}(k) = {\mathcal S}_{\mathrm{c} \gamma}(k) = {\mathcal S}_{\mathrm{c}\nu}(k).
\label{CDS0}
\end{equation}
In principle we might have the physical situation where ${\mathcal S}_{\mathrm{c}\gamma}(k)\neq 
{\mathcal S}_{\mathrm{c}\nu}(k)$. In the latter case, however, Eq. (\ref{NAD1}) 
implies that ${\mathcal S}_{\nu\gamma}(k) \neq 0$. This means we would not be dealing with a pure 
CDM-radiation mode but rather with a CDM-radiation mode supplemented by a neutrino-entropy mode 
(see also Eq. (\ref{NAD3})).
For the magnetized adiabatic mode the $\alpha$-expansion defined 
in Eq. (\ref{expans1}) only provides useful corrections to the leading order (constant) solution for $\xi(k,\tau)$. For the CDM-radiation solution $\xi$ is not constant but it rather vanishes as $\alpha$. 
Consequently, for $\alpha \ll 1$ and $k\tau \ll 1$, the asymptotic form of the solution is:
\begin{eqnarray}
&& h(k,\tau) = \overline{h}(k,\tau) = \frac{\sqrt{\alpha + 1}}{\alpha^2} \biggl[ 16 + \frac{2 (\alpha^2 - 4 \alpha - 8)}{\sqrt{\alpha + 1}}\biggr] {\mathcal D}_{*}(k)
\label{CDS1}\\
&& \xi(k,\tau) = - \frac{\overline{h}(k,\tau)}{6} - \frac{k^2 \tau_{1}^2}{108} \biggl(\frac{15 - 4 R_{\nu}}{15 + 2 R_{\nu}}\biggr) {\mathcal D}_{*}(k) x^3
\label{CDS2}\\
&& \delta_{\gamma}(k,\tau) = \frac{2}{3} \overline{h}(k,\tau) - R_{\gamma} \Omega_{\mathrm{B}}(k),
\label{CDS3}\\
&& \delta_{\nu}(k,\tau) = \frac{2}{3} \overline{h}(k,\tau) - R_{\gamma} \Omega_{\mathrm{B}}(k),
\label{CDS4}\\
&& \delta_{\mathrm{c}}(k,\tau) = - {\mathcal S}_{*}(k) + \frac{\overline{h}(k,\tau)}{2} - \frac{3}{4} R_{\gamma}\Omega_{\mathrm{B}}(k),
\label{CDS5}\\
&& \delta_{\mathrm{b}}(k,\tau)  = \frac{\overline{h}(k,\tau)}{2} - \frac{3}{4} R_{\gamma} \Omega_{\mathrm{B}}(k),
\label{CDS6}\\
&& \theta_{\gamma\mathrm{b}}(k,\tau) = \frac{k^2\tau}{4} [R_{\nu} \Omega_{\mathrm{B}}(k) - 4 \sigma_{\mathrm{B}}(k)]+  \frac{k^2 \tau_{1}}{6} {\mathcal D}_{*}(k) x^2,
\label{CDS7}\\
&& \theta_{\nu}(k,\tau) = \frac{k^2 \tau}{4} \biggl[ 4 \frac{R_{\gamma}}{R_{\nu}} \sigma_{\mathrm{B}}(k) -
R_{\gamma} \Omega_{\mathrm{B}}(k)\biggr] + \frac{k^2 \tau_{1}}{6} {\mathcal D}_{*}(k) x^2,
\label{CDS8}\\
&& \theta_{\mathrm{c}}(k,\tau) =0,
\label{CDS9}\\
&& \sigma_{\nu} (k,\tau)  = - \frac{R_{\gamma}}{R_{\nu}} \sigma_{\mathrm{B}}(k) + 
\frac{{\mathcal D}_{*}(k) k^2 \tau_{1}^2}{3 ( 15 + 2 R_{\nu})} x^3,
\label{CDS10}\\
&& {\mathcal F}_{\nu 3}(k, \tau)= \frac{8}{9}\frac{R_{\gamma}}{R_{\nu}}\biggl[ \sigma_{\mathrm{B}}(k)  - \frac{R_{\nu}}{4}\Omega_{\mathrm{B}}(k) \biggr] k\tau ,
\label{CDS11}
\end{eqnarray}
where, as established in Eq. (\ref{alpha}), $ x = \tau/\tau_{1}$ and 
${\mathcal D}_{*}(k)$  has been already defined in Eq. (\ref{Ddef}).
The function $\overline{h}(k,\tau)$ can be expanded for $\alpha \ll 1$:
\begin{equation}
\overline{h}(k,\tau) = \frac{\sqrt{\alpha + 1}}{\alpha^2} \biggl[ 16 + \frac{2 (\alpha^2 - 4 \alpha - 8)}{\sqrt{\alpha + 1}}\biggr] {\mathcal D}_{*}(k)=  \biggl( \alpha - \frac{5}{8} \alpha^2\biggr){\mathcal D}_{*}(k) + {\mathcal O}(\alpha^3).
\end{equation}
The CDM-radiation power spectra are assigned as ${\mathcal P}_{{\mathcal S}}(k) = {\mathcal A}_{{\mathcal S}} (k/k_{\mathrm{p}})^{n_{\mathrm{c}}-1}$ 
with the same pivot scale, i.e. $k_{\mathrm{p}} = 0.002\,\,\mathrm{Mpc}^{-1}$ adopted 
in the adiabatic case. In  Fig. \ref{Figure2}  the spectral 
index of each plot (i.e. $n_{\mathrm{c}}$) is specified in the 
corresponding legends\footnote{In the title of the plots the value of the optical depth 
has been also specified (i.e. $\tau=0.089$ as implied by the WMAP three year best fit). This notation is standard and we did not
deviate from it. However, it should be remarked that $\tau$ is consistently used in our script to 
denote the conformal time coordinate. This remark avoids any unwanted confusion.}.
To ease the comparison between Figs. \ref{Figure1} and  \ref{Figure2} the non-adiabatic 
amplitude is fixed to the same value of the adiabatic one, i.e. ${\mathcal A}_{{\mathcal S}} = {\mathcal A}_{{\mathcal R}} =2.35\times 10^{-9}$.
For $n_{\mathrm{c}} > 1$ a sort of hump arises in the Doppler region. In Fig. \ref{Figure2} 
(top-left plot)   the TT correlations are reported in the case $n_{\mathrm{c}} = 2$. 
The full line denotes in this plot the case $B_{\mathrm{L}}=0$ while the dashed 
and the dot dashed lines correspond to the case $B_{\mathrm{L}} = 10$ nG with 
two different magnetic spectral indices (as indicated in the legends). 
The top-right and bottom-left plots of Fig. \ref{Figure2} the EE and TE angular power spectra are illustrated 
with the same notations.
In the bottom-right plot the TT correlations are reported for three different values of the
non-adiabatic spectral index (i.e. $n_{\mathrm{c}} = 1,\,2,\,3$) when the parameters 
of the magnetized bckground are fixed. From the latter plot it is clear that, in the scale-invariant limit 
the hump is not present.

After careful numerical scrutiny of the magnetized CDM-radiation 
mode the following general aspects can be highlighted:
\begin{itemize}
\item{} the inclusion of the magnetic field 
diminishes the height of the hump of the TT angular power spectra and also suppresses 
the TE and EE correlations;
\item{} slight changes 
in the magnetic spectral index (e.g. from $1.2$ to $2$ in Fig. \ref{Figure2})  affect
quantitatively all the CMB observables;
\item{}  the phases of the EE angular power spectra (and, to a lesser degree, of the TE angular power spectra) are slightly shifted as $\ell >500$.
\end{itemize}
The TE correlation starts positive at large scales (see Fig. \ref{Figure2} bottom-left plot). 
This is in sharp contrast with what is experimentally 
observed (i.e. the TE correlations exhibit an anticorrelation peak \cite{WMAP1,WMAP2,WMAP3}). 
The  presence of the magnetic field can indeed reduce the TE amplitude but  
cannot change the sign of the first peak by making it negative. 
\renewcommand{\theequation}{5.\arabic{equation}}
\setcounter{equation}{0}
\section{Magnetized baryon-radiation mode}
\label{sec5}
If $\omega_{\mathrm{b}0} \ll \omega_{\mathrm{c}0}$ (as it happens in the $\Lambda$CDM paradigm) the baryon-radiation mode is, grossly 
speaking, less important than the CDM-radiation mode. 
The baryon-radiation mode touches upon the baryon-lepton fluid 
(which is also affected by the Ohmic current). Interesting  effects 
on the phases of the TE and EE angular power spectra are then potentially expected.
From Eq. (\ref{NAD2}) we will denote by $\overline{{\mathcal S}}_{*}(k)$ the common 
amplitude, in Fourier space, of the two relevant entropy fluctuations, i.e. 
\begin{equation}
\overline{{\mathcal S}}_{*}(k) = {\mathcal S}_{\mathrm{b} \gamma}(k) = {\mathcal S}_{\mathrm{b}\nu}(k).
\label{BS0}
\end{equation}
\begin{figure}[!ht]
\centering
\includegraphics[height=6.5cm]{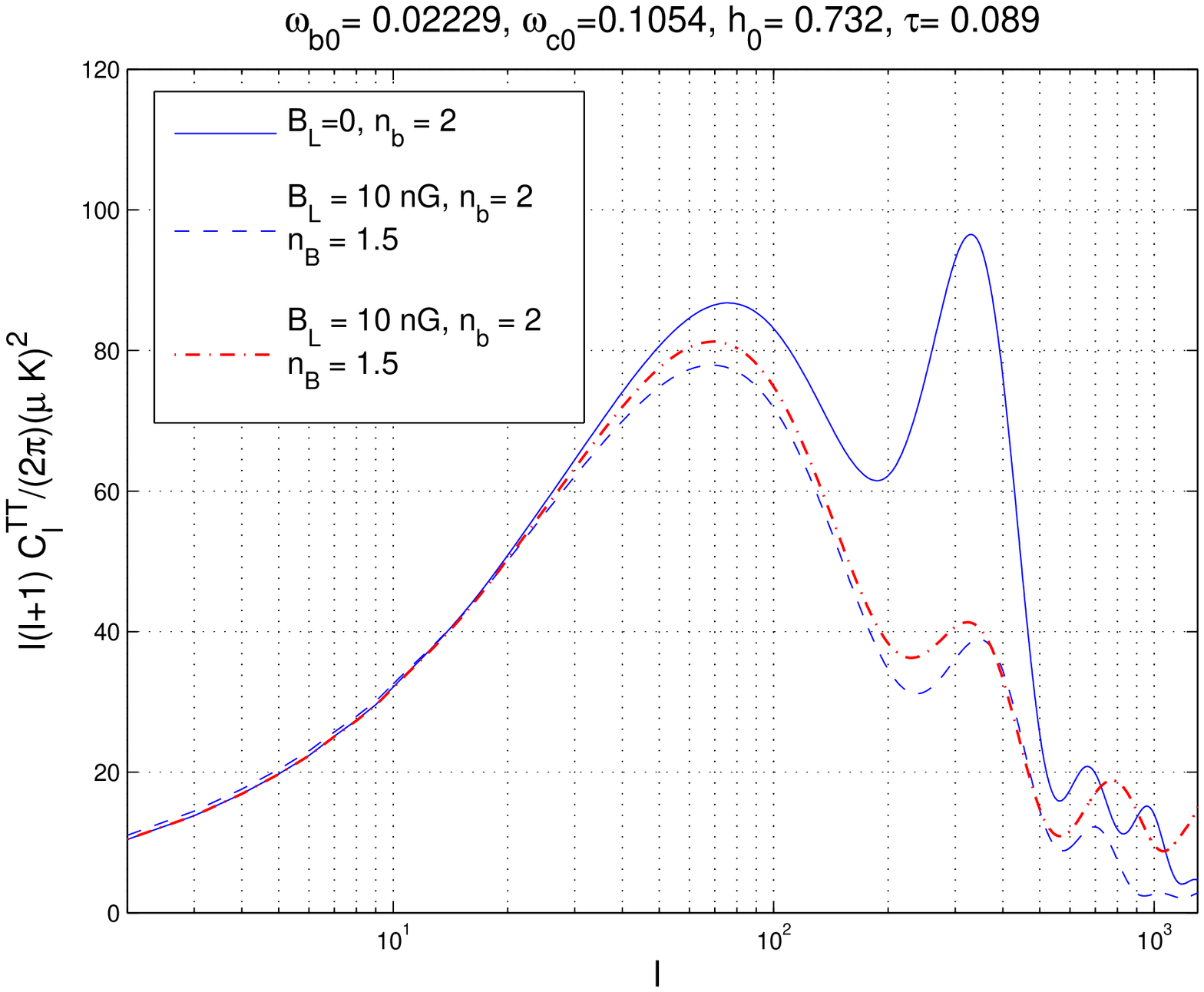}
\includegraphics[height=6.5cm]{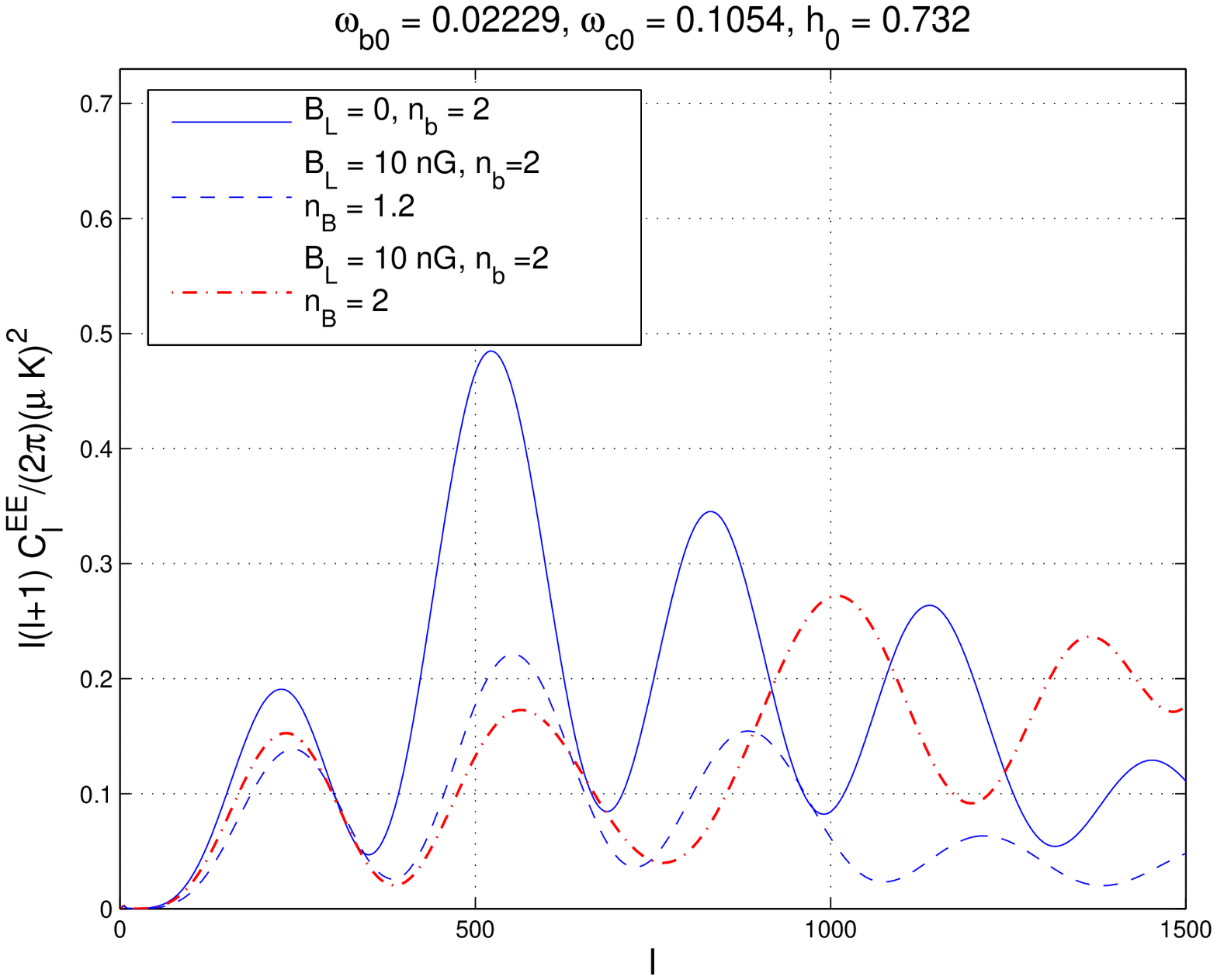}
\includegraphics[height=6.5cm]{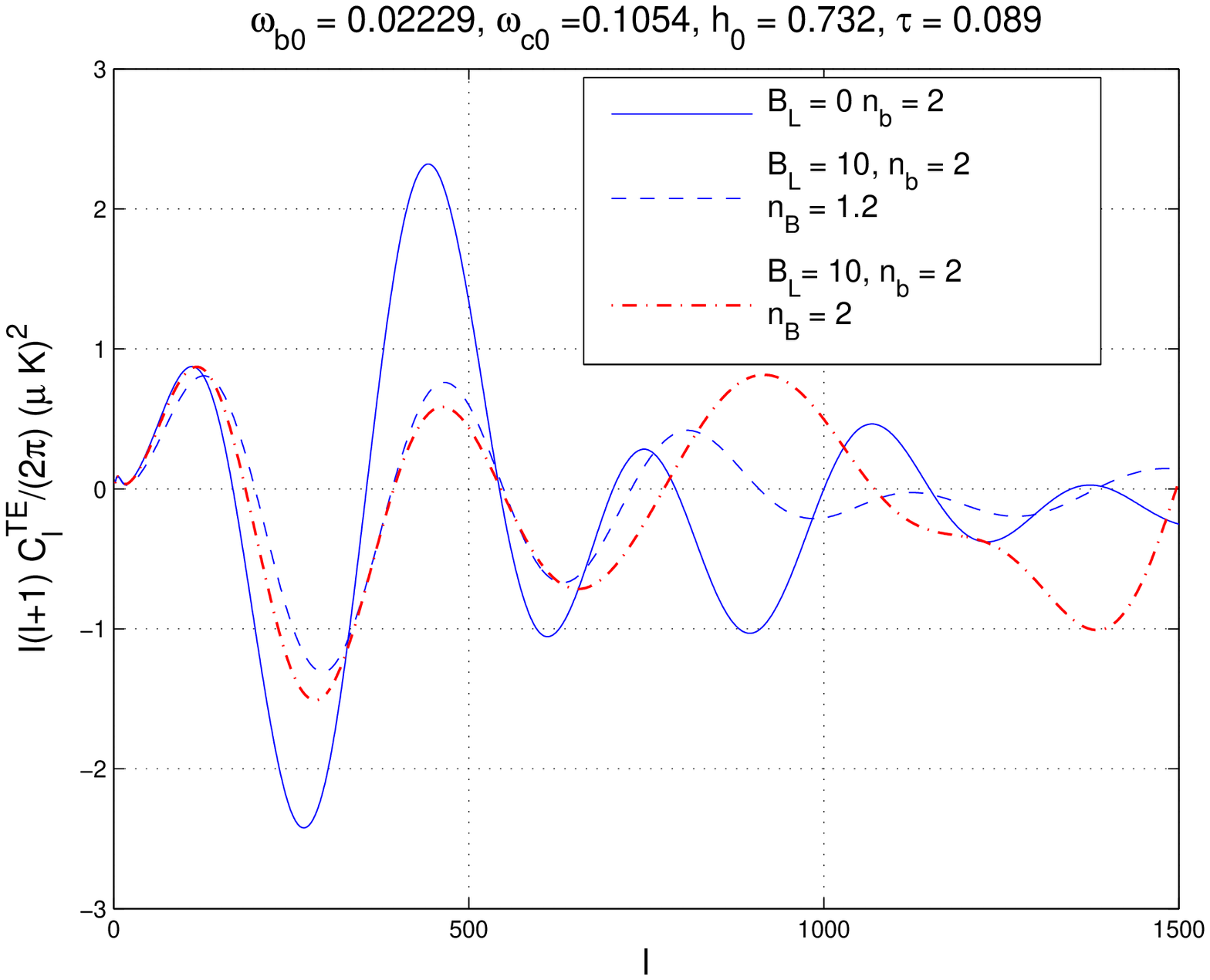}
\includegraphics[height=6.5cm]{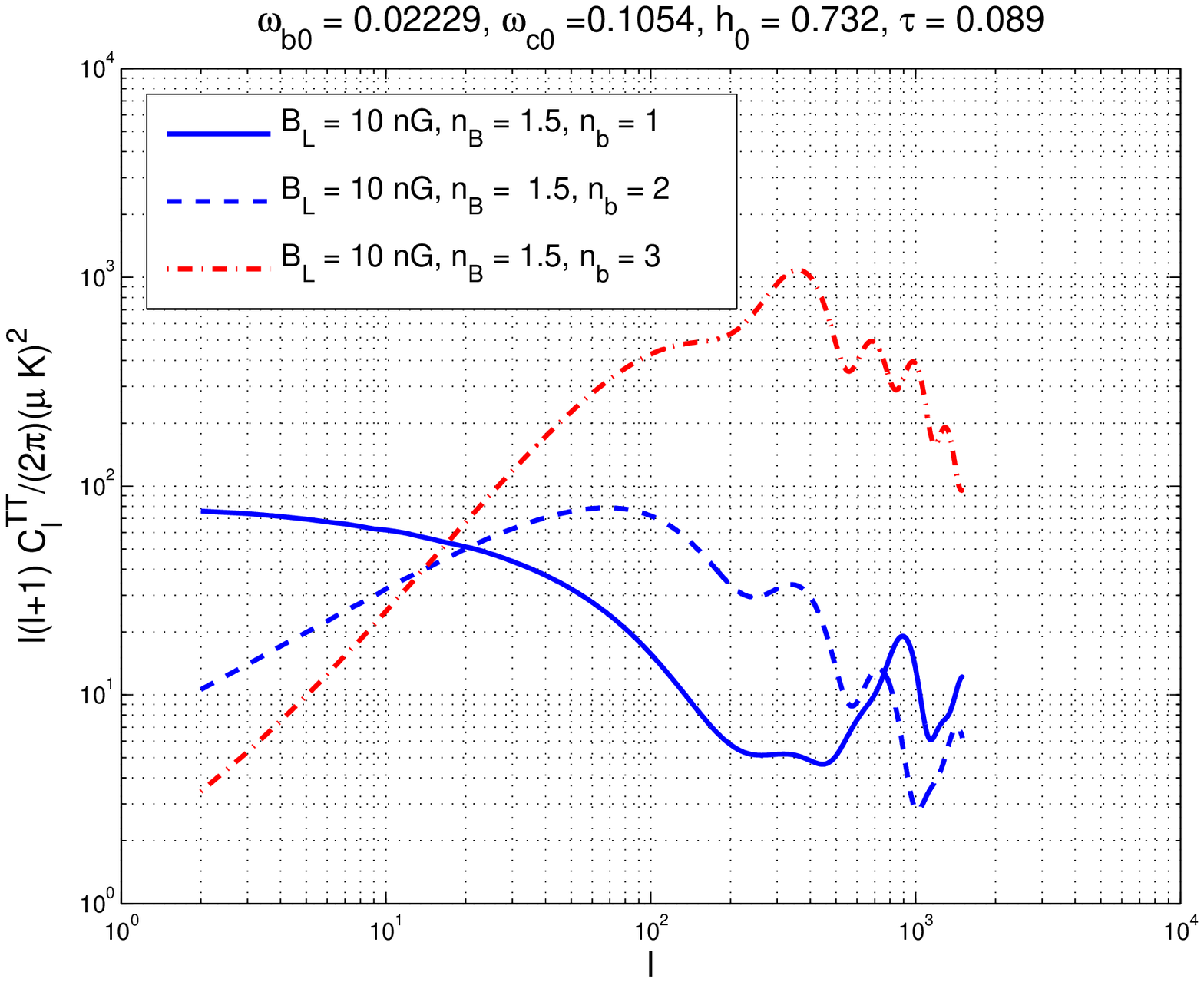}
\caption[a]{The magnetized baryon-radiation mode is illustrated for spatially flat models. To ease the comparison, the cosmological parameters are fixed to the same values of Figs. \ref{Figure1} and \ref{Figure2}.}
\label{Figure3}      
\end{figure}
For the reasons explained in the previous Section, ${\mathcal S}_{\mathrm{b}\gamma}(k)\neq 
{\mathcal S}_{\mathrm{b}\nu}(k)$ also demands  ${\mathcal S}_{\nu\gamma}(k) \neq 0$ and this 
would imply that the baryon-radiation mode is supplemented by a neutrino-entropy mode.
To make a distinction with respect to the 
CDM-radiation mode, the Fourier amplitude of the entropy fluctuation has been denoted, in the baryon-radiation case, as $\overline{{\mathcal S}}_{*}(k)$.
Direct integration of the truncated Einstein-Boltzmann hierarchy 
summarized in Appendix \ref{APPA} leads to the following initial conditions:
\begin{eqnarray}
&& h(k,\tau) = \overline{h}(k,\tau)= \biggl( \alpha - \frac{5}{8} \alpha^2\biggr) \overline{{\mathcal D}}_{*}(k)
\label{BS1}\\
&& \xi(k,\tau) = - \frac{\overline{h}(k,\tau)}{6} - \frac{k^2 \tau_{1}^2}{108} \biggl(\frac{15 - 4 R_{\nu}}{15 + 2 R_{\nu}}\biggr) \overline{{\mathcal D}}_{*}(k) x^3
\label{BS2}\\
&& \delta_{\gamma}(k,\tau) = \frac{2}{3} \overline{h}(k,\tau) - R_{\gamma} \Omega_{\mathrm{B}}(k),
\label{BS3}\\
&& \delta_{\nu}(k,\tau) = \frac{2}{3} \overline{h}(k,\tau) - R_{\gamma} \Omega_{\mathrm{B}}(k),
\label{BS4}\\
&& \delta_{\mathrm{c}}(k,\tau) =  \frac{\overline{h}(k,\tau)}{2} - \frac{3}{4} R_{\gamma}\Omega_{\mathrm{B}}(k),
\label{BS5}\\
&& \delta_{\mathrm{b}}(k,\tau)  = - \overline{{\mathcal S}}_{*}(k) +\frac{\overline{h}(k,\tau)}{2} - \frac{3}{4} R_{\gamma} \Omega_{\mathrm{B}}(k),
\label{BS6}\\
&& \theta_{\gamma\mathrm{b}}(k,\tau) = \frac{k^2\tau}{4} [R_{\nu} \Omega_{\mathrm{B}}(k) - 4 \sigma_{\mathrm{B}}(k)] +  \frac{k^2 \tau_{1}}{6} \overline{{\mathcal D}}_{*}(k)x^2,
\label{BS7}\\
&& \theta_{\nu}(k,\tau) = \frac{k^2 \tau}{4} \biggl[ 4 \frac{R_{\gamma}}{R_{\nu}} \sigma_{\mathrm{B}}(k) -
R_{\gamma} \Omega_{\mathrm{B}}(k)\biggr] + \frac{k^2 \tau_{1}}{6} \overline{{\mathcal D}}_{*}(k) x^2,
\label{BS8}\\
&& \theta_{\mathrm{c}}(k,\tau) =0,
\label{BS9}\\
&& \sigma_{\nu} (k,\tau)  = - \frac{R_{\gamma}}{R_{\nu}} \sigma_{\mathrm{B}}(k) + 
\frac{\overline{{\mathcal D}}_{*}(k) k^2 \tau_{1}^2}{3( 15 + 2 R_{\nu})} x^3,
\label{BS10}\\
&& {\mathcal F}_{\nu 3}(k, \tau)= \frac{8}{9}\frac{R_{\gamma}}{R_{\nu}} k\tau \biggl[  \sigma_{\mathrm{B}}(k)  - \frac{R_{\nu}}{4}\Omega_{\mathrm{B}}(k)\biggr],
\label{BS11}
\end{eqnarray}
where, following the analog notation used in Eq. (\ref{Ddef})  for the CDM-radiation mode,
\begin{equation}
\overline{{\mathcal D}}_{*}(k) = \biggl[ \frac{\omega_{\mathrm{b}0}}{\omega_{\mathrm{M}0}} \overline{{\mathcal S}}_{*}(k) + 
\frac{3}{4} R_{\gamma} \Omega_{\mathrm{B}}(k)  \biggr],
\qquad 
\frac{\omega_{\mathrm{b}0}}{\omega_{\mathrm{M}0}} = \frac{\omega_{\mathrm{b}0}}{\omega_{\mathrm{c}0} + \omega_{\mathrm{b}0}}.
\end{equation}
The power spectrum of the baryon-radiation mode is assigned 
as ${\mathcal P}_{\overline{{\mathcal S}}}(k) = {\mathcal A}_{\overline{{\mathcal S}}}(k/k_{\mathrm{p}})^{n_{\mathrm{b}} -1}$ where $n_{\mathrm{b}}$ is the corresponding spectral index.
Comparing Eqs. (\ref{BS1})--(\ref{BS11})  with Eqs. (\ref{CDS1})--(\ref{CDS11}), there 
are formal analogies but also some differences. In the case of Eqs. (\ref{BS1})--(\ref{BS11}) 
the baryons are directly affected and, therefore,
\begin{itemize}
\item{} as anticipated, for the fiducial set of parameters of the $\Lambda$CDM model 
the amplitude of the TT, EE and TE angular power spectra will 
be comparatively smaller than in the case of the CDM-radiaation mode;
\item{} a stronger effect on the phases is expected when the magnetic field is consistently included in the calculation.
\end{itemize}
In Fig. \ref{Figure3} the TT, EE and TE angular power spectra  are 
illustrated when the amplitude is selected in such a way that ${\mathcal A}_{\overline{{\mathcal S}}} = {\mathcal A}_{{\mathcal R}}$. The comparison 
of Fig. \ref{Figure3} and \ref{Figure2} suggests that, indeed 
the magnetized CDM-radiation mode is very similar to the 
magnetized baryon-radiation mode.  Also in the case of the baryon 
radiation-mode the hump arises for $n_{\mathrm{b}}> 1$
The solution of the Hamiltonian and of the momentum constraints implies 
a definite relative sign of the magnetic and of the entropic 
contributions. As a consequence the amplitude of the angular power spectra diminishes with 
respect to the case $B_{\mathrm{L}}=0$. 
The phases of oscillations of the TE and EE angular power spectra 
(see top-right and bottom-left plots in Fig. \ref{Figure3}) are shifted as the magnetic field strength increases.

 \renewcommand{\theequation}{6.\arabic{equation}}
\setcounter{equation}{0}
\section{Magnetized neutrino-entropy mode}
\label{sec6}
In the case of the magnetized neutrino-entropy mode the non-adiabatic fluctuation 
belongs to the neutrino sector, i.e. 
\begin{equation}
\tilde{{\mathcal S}}_{*}(k) = \frac{3}{4} [\delta_{\gamma}(k,\tau) - \delta_{\nu}(k\tau)].
\end{equation}
The lowest order form of the solution can be obtained by setting $\xi(k,\tau) = h(k,\tau) =0$. This aspect already highlights the difference 
of the neutrino-entropy mode in comparison with both,
the previous non-adiabatic modes and 
with the adiabatic solution. In the case of the adiabatic solution, to lowest 
order in the expansion of Eq. (\ref{expans1}), the curvature perturbations 
are constant in time, i.e. 
\begin{equation}
{\mathcal R}(k,\alpha) \simeq {\mathcal R}_{*}(k) + {\mathcal O}(k^2\tau^2) + {\mathcal O}(\alpha).
\label{comp1}
\end{equation}
\begin{figure}[!ht]
\centering
\includegraphics[height=6.5cm]{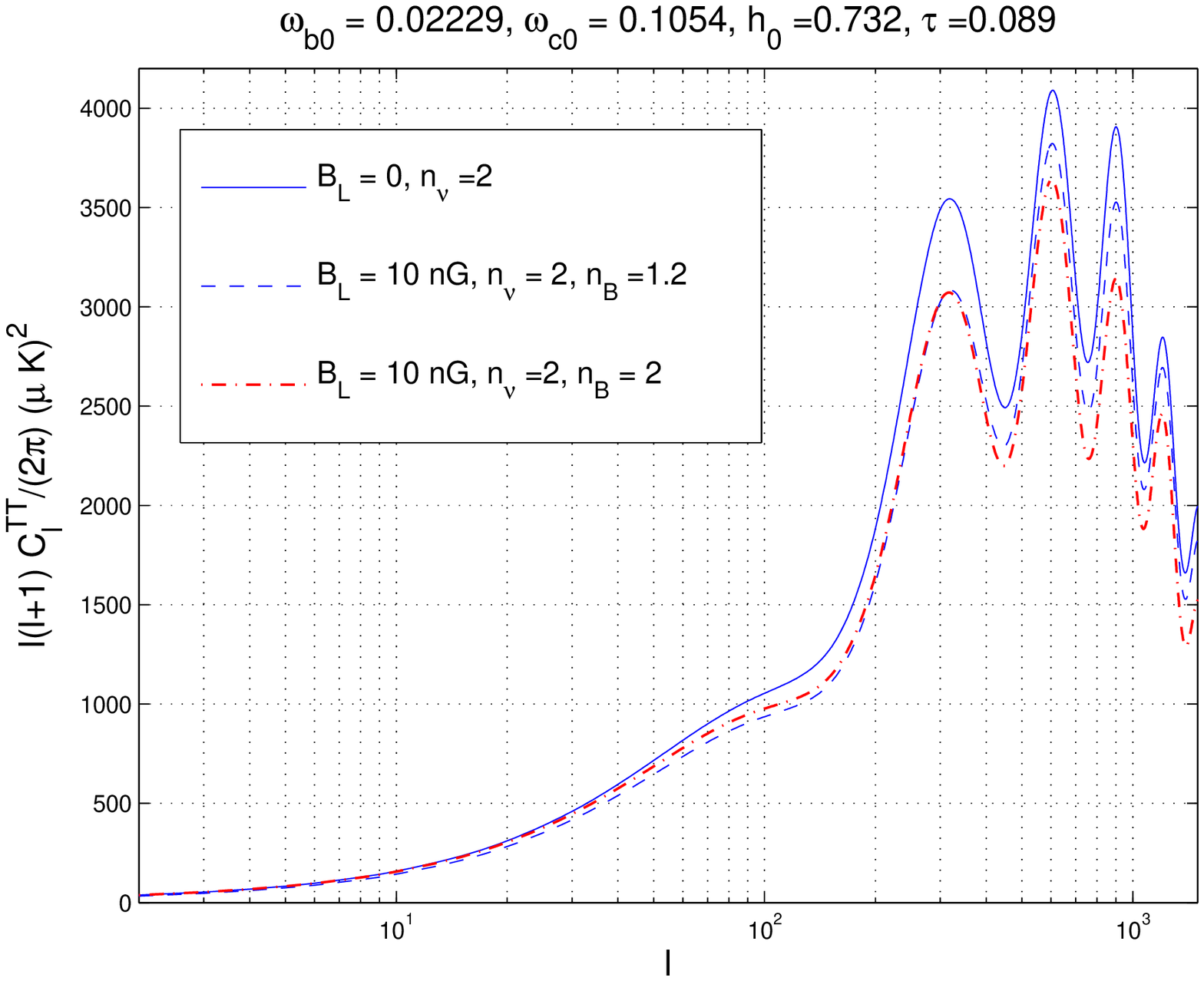}
\includegraphics[height=6.5cm]{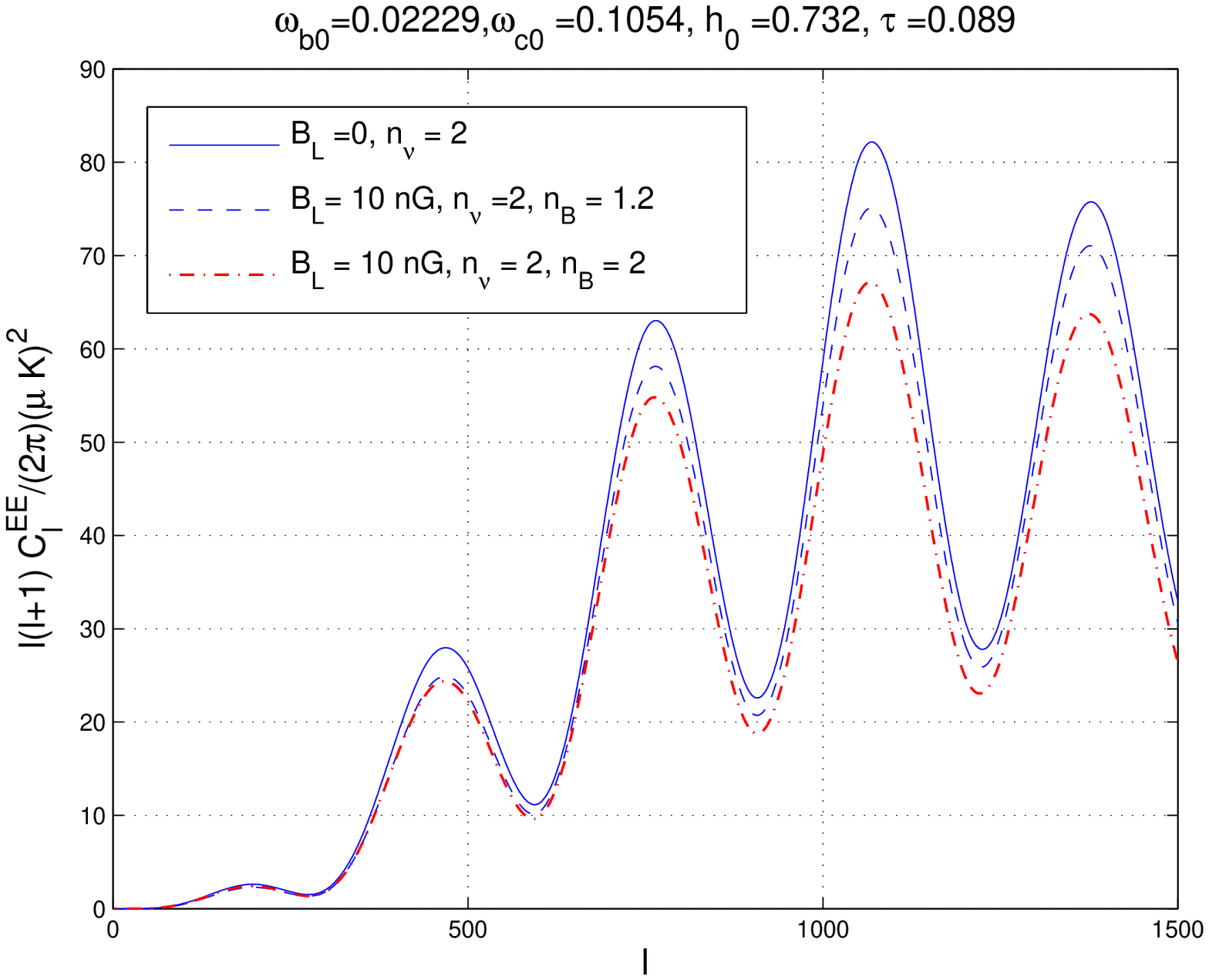}
\includegraphics[height=6.5cm]{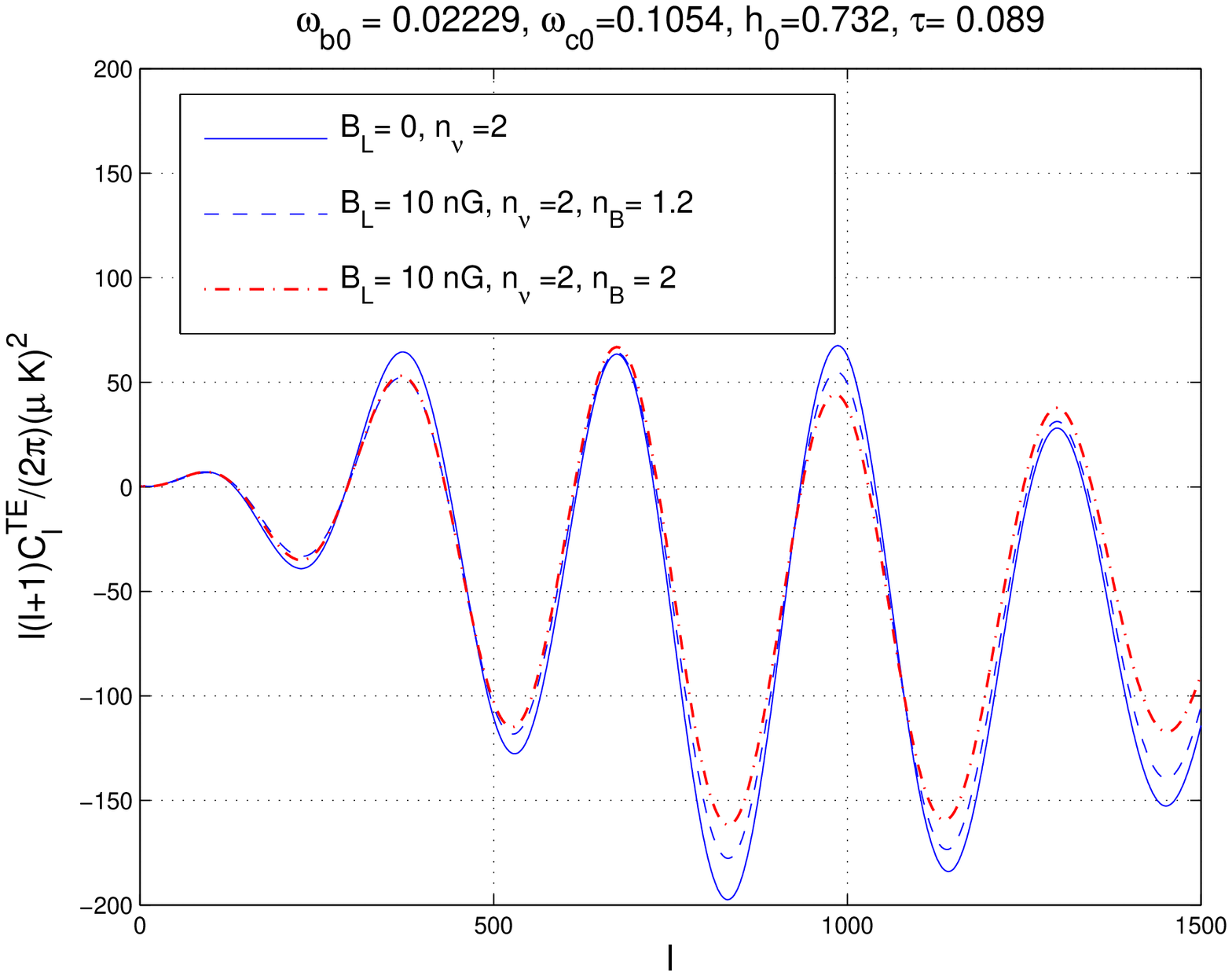}
\includegraphics[height=6.5cm]{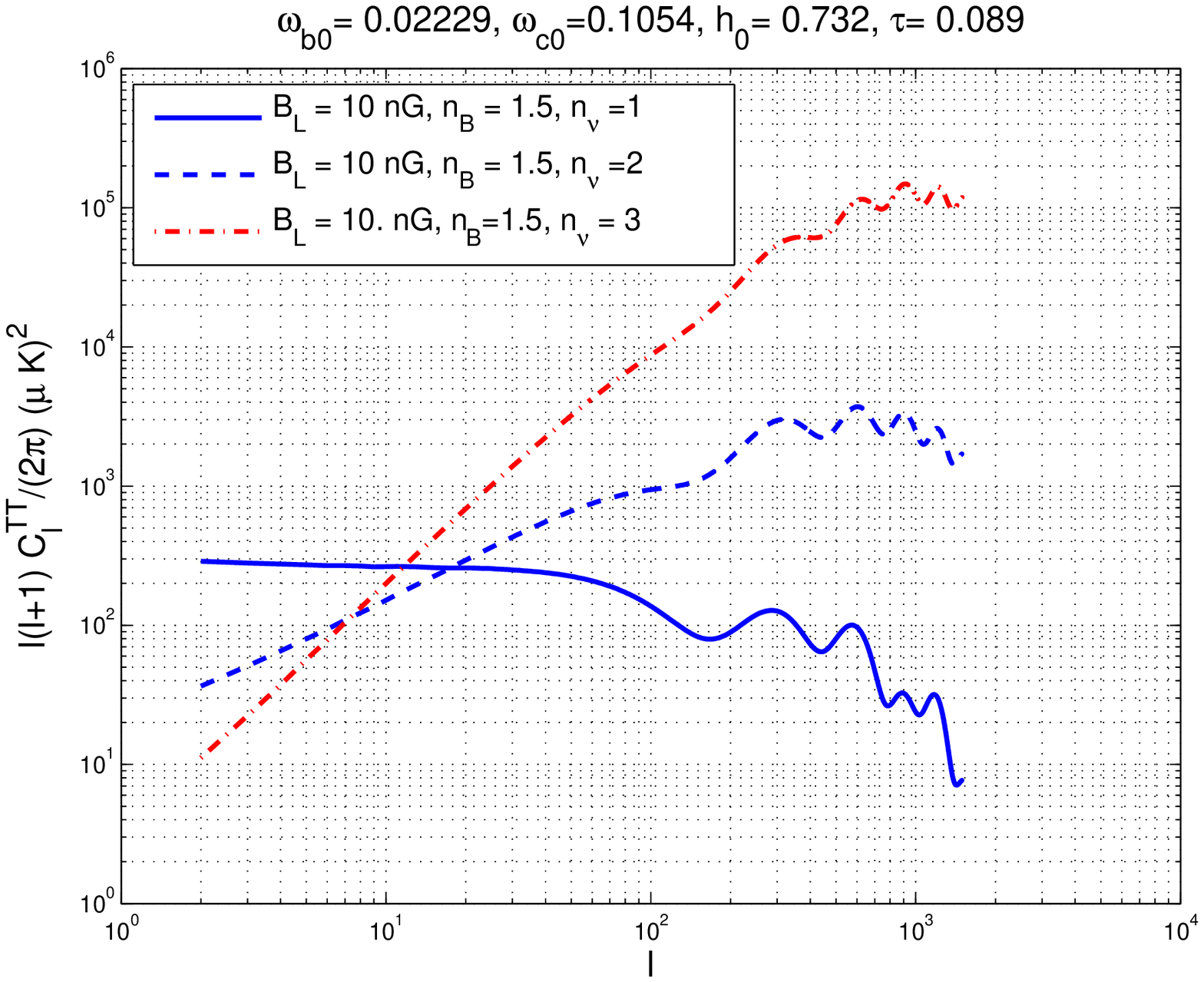}
\caption[a]{The angular power spectra obtained in the case 
of the neutrino-entropy mode are illustrated for the same 
set of cosmological parameters used in Figs. \ref{Figure1}, \ref{Figure2}, 
and \ref{Figure3}.}
\label{Figure4}      
\end{figure}
For the CDM and baryon-radiation modes the 
curvature perturbations vanish as $\alpha$, to lowest order in $k\tau$, i.e.
${\mathcal R} \propto \alpha + {\mathcal O}(k^2\tau^2)$. In the 
case of the neutrino-entropy mode the behaviour is intermediate 
since ${\mathcal R}(k,\alpha) \simeq {\mathcal O}(k^2\tau^2)$.
The specific form of the neutrino-entropy mode is then obtained by integration 
of the truncated Einstein-Boltzmann hierarchy reported in Appendix \ref{APPA}. The final result is:
\begin{eqnarray}
&& h(k,\tau) = \frac{\omega_{\mathrm{b}0}}{\omega_{\mathrm{M}0}}\biggl[ \frac{ R_{\nu} }{3}
\tilde{{\mathcal S}}_{*}(k) -  \sigma_{\mathrm{B}}(k) + \frac{R_{\nu}}{4}\Omega_{\mathrm{B}}(k)\biggr] k^2 \tau_{1}^2 x^{3},
\label{NS1}\\
&& \xi(k,\tau) =-\frac{2 R_{\gamma} R_{\nu}}{9( 4 R_{\nu} + 15)} \biggl[ \tilde{{\mathcal S}}_{*}(k) - 
\frac{3}{R_{\nu}} \sigma_{\mathrm{B}}(k) + \frac{3}{4} \Omega_{\mathrm{B}}(k) \biggr] k^2 \tau^2,
\label{NS2}\\
&& \delta_{\gamma}(k,\tau) = \frac{4}{3} \tilde{{\mathcal S}}_{*}(k) R_{\nu} - R_{\gamma}
\Omega_{\mathrm{B}}(k,\tau)   -\frac{1}{6} \biggl[ \frac{4}{3} R_{\nu} {{\mathcal S}}_{*}(k) - 4 \sigma_{\mathrm{B}}(k) + R_{\nu} \Omega_{\mathrm{B}}(k)\biggr] k^2 \tau^2,
\label{NS3}\\
&& \delta_{\nu}(k,\tau) = - \frac{4}{3} \tilde{{\mathcal S}}_{*}(k) R_{\gamma} - 
R_{\gamma}\Omega_{\mathrm{B}}(k) - \frac{1}{6}\biggl[ - \frac{4}{3} R_{\gamma} \tilde{{\mathcal S}}_{*}(k) + 4 \frac{R_{\gamma}}{R_{\nu}} \sigma_{\mathrm{B}}(k) - R_{\gamma}\Omega_{\mathrm{B}}(k) \biggr] k^2 \tau^2,
\label{NS4}\\
&& \delta_{\mathrm{c}}(k,\tau) =   
\frac{1}{2} \frac{\omega_{\mathrm{b}0}}{\omega_{\mathrm{M}0}}\biggl[ \frac{ R_{\nu} }{3}
\tilde{{\mathcal S}}_{*}(k) -  \sigma_{\mathrm{B}}(k) + \frac{R_{\nu}}{4}\Omega_{\mathrm{B}}(k)\biggr] k^2 \tau_{1}^2 x^{3}
\label{NS5}\\
&& \delta_{\mathrm{b}}(k,\tau) = - \frac{1}{2} \biggl[ \frac{R_{\nu}}{3} \tilde{{\mathcal S}}_{*}(k) -  \sigma_{\mathrm{B}}(k) + \frac{R_{\nu}}{4} \Omega_{\mathrm{B}}(k) \biggr] k^2 \tau^2,
\label{NS6}\\
&& \theta_{\gamma\mathrm{b}}(k,\tau) = \biggl[ \frac{R_{\nu}}{3} \tilde{{\mathcal S}}_{*}(k) - 
 \sigma_{\mathrm{B}}(k) + \frac{R_{\nu}}{4}\Omega_{\mathrm{B}}(k)\biggr] k^2 \tau - \frac{3}{4} \frac{\omega_{\mathrm{b}0}}{\omega_{\mathrm{M}0}} \biggl[ \frac{R_{\nu}}{3 R_{\gamma}} \tilde{{\mathcal S}}_{*} - \frac{\sigma_{\mathrm{B}}}{R_{\gamma}} + \frac{R_{\nu}}{4 R_{\gamma}} \Omega_{\mathrm{B}}\biggr] k^2\tau_{1} x^2,
\label{NS7}\\
&& \theta_{\nu}(k,\tau) = \biggl[ - \frac{R_{\gamma}}{3} \tilde{{\mathcal S}}_{*}(k) + 
 \frac{R_{\gamma}}{R_{\nu}} \sigma_{\mathrm{B}}(k) - \frac{R_{\gamma}}{4} \Omega_{\mathrm{B}}(k)\biggr] k^2 \tau,
\label{NS8}\\
&& \theta_{\mathrm{c}} =0,
\label{NS9}\\
&& \sigma_{\nu}(k,\tau) = - \frac{R_{\gamma}}{R_{\nu}} \sigma_{\mathrm{B}}(k) - \frac{2 R_{\gamma}}{3(4 R_{\nu} + 15)} \biggl[\tilde{{\mathcal S}}_{*}(k) - \frac{3}{R_{\nu}} \sigma_{\mathrm{B}}(k) + \frac{3}{4} \Omega_{\mathrm{B}}(k)\biggr] k^2 \tau^2.
\label{NS10}
\end{eqnarray}
In the case of the neutrino-entropy mode the power spectra are assigned 
as ${\mathcal P}_{\tilde{{\mathcal S}}} = {\mathcal A}_{\tilde{{\mathcal S}}} (k/k_{\mathrm{p}})^{(n_{\nu} -1)}$
where $n_{\nu}$ is the corresponding spectral index.
In Fig. \ref{Figure4} the numerical integration of the angular power spectra is reported for 
${\mathcal A}_{\tilde{{\mathcal S}}} = {\mathcal A}_{{\mathcal R}} = 2.35 \times 10^{-9}$. 
Figure \ref{Figure4} also shows two interesting features:
\begin{itemize}
\item{} the hump disappears not only in the scale-invariant limit (i.e. 
$n_{\nu} =1$) but also for $n_{\nu} >1$; in the latter cases, however, 
the amplitude increases;
\item{} as in the other two cases the Hamiltonian and momentum constraints 
imply that the inclusion of the magnetic field always tend to 
diminish the heights of the acoustic oscillations.
\end{itemize}
 As it is typical also of the other modes the TE angular 
 power spectrum always exhibits a correlation 
 (rather than anticorrelation) at large-scales.
 The phase shifts in the polarization observables do not arise 
for the neutrino-entropy mode: the magnetic and 
the entropic contributions do not interfere directly since 
the neutrinos, unlike the baryons and electrons, only feel 
the magnetic field through the constraints and through the ansiotropic stress.

For sake of completeness, the solution of the truncated Einstein-Boltzmann hierarchy will be derived for the neutrino-velocity mode.
The neutrino-velocity mode is rather peculiar since it is, strictly speaking, not an entropic mode. It is derived by requiring that 
the initial fluctuation resides in the velocity of the neutrinos.
The momentum constraint stemming from the $(0i)$ component 
of the perturbed Einstein equations (i.e. Eq. (\ref{MOM1})) then demands 
that also the photon-baryon velocity is perturbed at early 
times so that, overall, the momentum constraint is 
satisfied at the onset of the numerical integration. From this 
observation the leading order solution implies that $h\simeq 0$ and $\xi \simeq 0$ and, therefore, 
\begin{equation}
R_{\nu} \theta_{\nu} + R_{\gamma}( 1 + R_{\mathrm{b}}) \theta_{\gamma\mathrm{b}} \to 0,
\end{equation}
at early times. Of course, both $\theta_{\nu}$ and $\theta_{\gamma\mathrm{b}}$ are both non-vanishing.
The neutrino-velocity mode can be initialized by setting:
\begin{eqnarray}
&& h(k,\tau) = - \frac{3}{2}  \tilde{{\mathcal S}}_{*}(k)\frac{R_{\nu}}{R_{\gamma}}
\frac{\omega_{\mathrm{b}0}}{\omega_{\mathrm{M}0}}k\tau_{1}x^2 ,
\label{NV1}\\
&& \xi(k,\tau) = \frac{4}{3} \frac{R_{\nu}\tilde{{\mathcal S}}_{*}(k)}{4 R_{\nu} + 5}  k\tau,
\label{NV2}\\
&& \delta_{\gamma}(k,\tau) = \frac{4}{3} \frac{R_{\nu}}{R_{\gamma}} \tilde{{\mathcal S}}_{*}(k) k\tau - \tilde{{\mathcal S}}_{*}(k)  \frac{R_{\nu}(R_{\gamma}+1)}{R_{\gamma}^2}
\frac{\omega_{\mathrm{b}0}}{\omega_{\mathrm{M}0}}k\tau_{1} x^2- R_{\gamma} \Omega_{\mathrm{B}},
\label{NV3}\\
&& \delta_{\nu}(k,\tau) = - \frac{4}{3} \tilde{{\mathcal S}}_{*}(k) k\tau - \tilde{{\mathcal S}}_{*}(k) \frac{R_{\nu}}{R_{\gamma}}
\frac{\omega_{\mathrm{b}0}}{\omega_{\mathrm{M}0}}  k\tau_{1} x^2- R_{\gamma} \Omega_{\mathrm{B}}
\label{NV4}\\
&& \delta_{\mathrm{c}}(k,\tau) =   - \frac{3}{4}  \tilde{{\mathcal S}}_{*}(k) \frac{R_{\nu}}{R_{\gamma}}
\frac{\omega_{\mathrm{b}0}}{\omega_{\mathrm{M}0}}k\tau_{1} x^2
\label{NV5}\\
&& \delta_{\mathrm{b}}(k,\tau) = \frac{R_{\nu}}{R_{\gamma}} \tilde{{\mathcal S}}_{*}(k) k\tau  - \frac{3}{4} \tilde{{\mathcal S}}_{*}(k) \frac{R_{\nu}(R_{\gamma} +1)}{R_{\gamma}^2}
\frac{\omega_{\mathrm{b}0}}{\omega_{\mathrm{M}0}} k\tau_{1} x^2,
\label{NV6}\\
&& \theta_{\gamma\mathrm{b}}(k,\tau) = - \frac{R_{\nu}}{R_{\gamma}} k \tilde{{\mathcal S}}_{*}(k) + 
\frac{3}{2} \biggl(\frac{\omega_{\mathrm{b}0}}{\omega_{\mathrm{M}0}}\biggr) \frac{R_{\nu}}{R_{\gamma}^2} 
\tilde{{\mathcal S}}_{*}(k)  k x - \frac{9}{4} \frac{R_{\nu}}{R_{\gamma}^3} \biggl(\frac{\omega_{\mathrm{b}0}}{\omega_{\mathrm{M}0}}\biggr)^2 \tilde{{\mathcal S}}_{*}(k) k x^2 
\nonumber\\
&& - \frac{3 R_{\nu}}{8 R_{\gamma}^2} \frac{\omega_{\mathrm{b}0}}{\omega_{\mathrm{M}0}} k \tilde{{\mathcal S}}_{*}x^2+ \frac{R_{\nu}}{6 R_{\gamma}} \tilde{{\mathcal S}}_{*}k^3 \tau^2+ \frac{k^2 \tau}{4} ( R_{\nu} \Omega_{\mathrm{B}} 
- 4\sigma_{\mathrm{B}}),
\label{NV8}\\
&& \theta_{\nu}(k,\tau) = k \tilde{{\mathcal S}}_{*}(k) - \frac{4 R_{\nu} + 9}{6 (4 R_{\nu} + 5)} \tilde{{\mathcal S}}_{*}(k) k^3 \tau^2 - \frac{R_{\gamma}}{4}\biggl[ \Omega_{\mathrm{B}} - \frac{4}{R_{\nu}} \sigma_{\mathrm{B}}\biggr] k^2 \tau
\label{NV7}\\
&& \theta_{\mathrm{c}}=0,
\label{NV6a}\\
&& \sigma_{\nu}(k,\tau) = - \frac{R_{\gamma}}{R_{\nu}} \sigma_{\mathrm{B}} +  \frac{4 \tilde{{\mathcal S}}_{*}(k)}{4 R_{\nu} + 5} k \tau,
\label{NV9}
\end{eqnarray}
where, as established in Eq. (\ref{alpha}), $x = \tau/\tau_{1}$. 
The quantity $\tilde{{\mathcal S}}_{*}(k)$ arising in the neutrino-velocity 
mode is physically different from the analog variable appearing 
in the solution of the neutrino-entropy mode. Here it measures
the spectrum of $\theta_{\nu}$ (or $\theta_{\gamma\mathrm{b}}$).
While the usual non-adiabatic modes 
can be described also in different gauges, the neutrino-velocity 
mode might even diverge if we move from the synchronous gauge.
In the longitudinal gauge
the scalar fluctuations of the geometry are defined, in the spatially 
flat case, as 
$\delta_{\mathrm{s}} g_{00} = 2 a^2 \phi$ and as $\delta_{\mathrm{s}} g_{ij} = 2 a^2 \psi \delta_{ij}$. The relation between the longitudinal 
and the synchronous degrees of freedom is given by 
\begin{equation}
\phi = - \frac{[(h + 6 \xi)'' + {\mathcal H} (h + 6 \xi)']}{2 k^2},\qquad 
\psi = - \xi +  \frac{{\mathcal H} (h + 6 \xi)'}{2 k^2}.
\label{tr1}
\end{equation}
From Eqs. (\ref{NS1}) and (\ref{NS2}) 
the values of $\phi$ and $\psi$ can then be
obtained to leading order in $k\tau$ and they are:
\begin{equation}
\phi(k) = \frac{8 R_{\gamma} R_{\nu}}{3 ( 4 R_{\nu} + 15)} \biggl[ \tilde{{\mathcal S}}_{*}(k) - 
\frac{3}{R_{\nu}} \sigma_{\mathrm{B}}(k) + \frac{3}{4} \Omega_{\mathrm{B}}(k) \biggr],\qquad \psi(k) = - \frac{\phi(k)}{2}.
\label{tr2}
\end{equation}
The same exercise can be performed in the case of the neutrino-velocity mode. Thus, inserting Eqs. (\ref{NV1}) and (\ref{NV2}) 
into Eq. (\ref{tr1}) we do get that 
\begin{equation}
\phi(k,\tau) = - \frac{4 R_{\nu} \tilde{{\mathcal S}}_{*}(k)}{(4 R_{\nu} + 5)|k\tau|}, \qquad \psi(k,\tau) = - \phi(k,\tau).
\label{tr3}
\end{equation}
Consequently, the neutrino-velocity mode is singular in the 
longitudinal gauge. In the same gauge the neutrino-entropy mode is not singular. It is therefore mandatory to use the synchronous frame to describe the neutrino-velocity mode. The unphysical nature 
of this divergence is clear since the gauge-invaraint curvature 
perturbations, i.e. ${\mathcal R}$, is regular for both modes.
\renewcommand{\theequation}{7.\arabic{equation}}
\setcounter{equation}{0}
\section{Mixtures of magnetized initial conditions}
\label{sec7}
Assuming a precise knowledge of the thermal history 
of the Universe from the Planck energy scale 
down to photon decoupling, then,  definite 
predictions on the CMB observables can be made. 
To make the predictions more definite, it is wise
to assume that the field content during inflation 
is rather limited. If we just demand a single 
adiabatic mode, as in the case of the $\Lambda$CDM 
paradigm, then a single inflaton field must be present.
This simple model is constrained by experimental 
data. More complicated models will be more constrained as soon as 
more accurate data will be available \cite{planck}. 
When magnetic fields are consistently 
included in the pre-equality physics,
the issue of initial conditions becomes particularly important 
and it has been recently shown that different thermal histories 
can imply slightly different amplitudes of the curvature 
perturbations induced by a magnetized background \cite{mg4}.

In a pragmatic perspective it is  plausible 
to start confronting with observations the simplest 
model. In the case of \cite{giokunz1}, the m$\Lambda$CDM 
scenario is the simplest option since a magnetized adiabatic mode is 
assumed and the magnetic fields introduce 
only two new parameters in comparison with the $\Lambda$CDM 
paradigm. It is however equally plausible 
to bear in mind also different possibilities.

The most general initial conditions of the magnetized 
CMB anisotropies will be given, presumably, as a mixture 
of various modes. As previously observed the TT angular power spectra will be typically distorted and shifted upwards if only the magnetized  adiabatic mode is assumed in the pre-equality plasma.
The magnetized 
CDM-radiation mode tends to reduce the amplitude of the TT 
power spectra in comparison 
with the case when the magnetic fields are absent. 
Since the magnetized adiabatic mode and the CDM-radiation 
mode are both solutions 
of the truncated Einstein-Boltzmann hierarchy, also 
their combination will lead to a viable initial condition.
The same reasoning can be applied also in more complicated cases, 
like the one where, on top of the magnetized adiabatic modes 
there are two or three non-adiabatic modes. 
\begin{figure}[!ht]
\centering
\includegraphics[height=7cm]{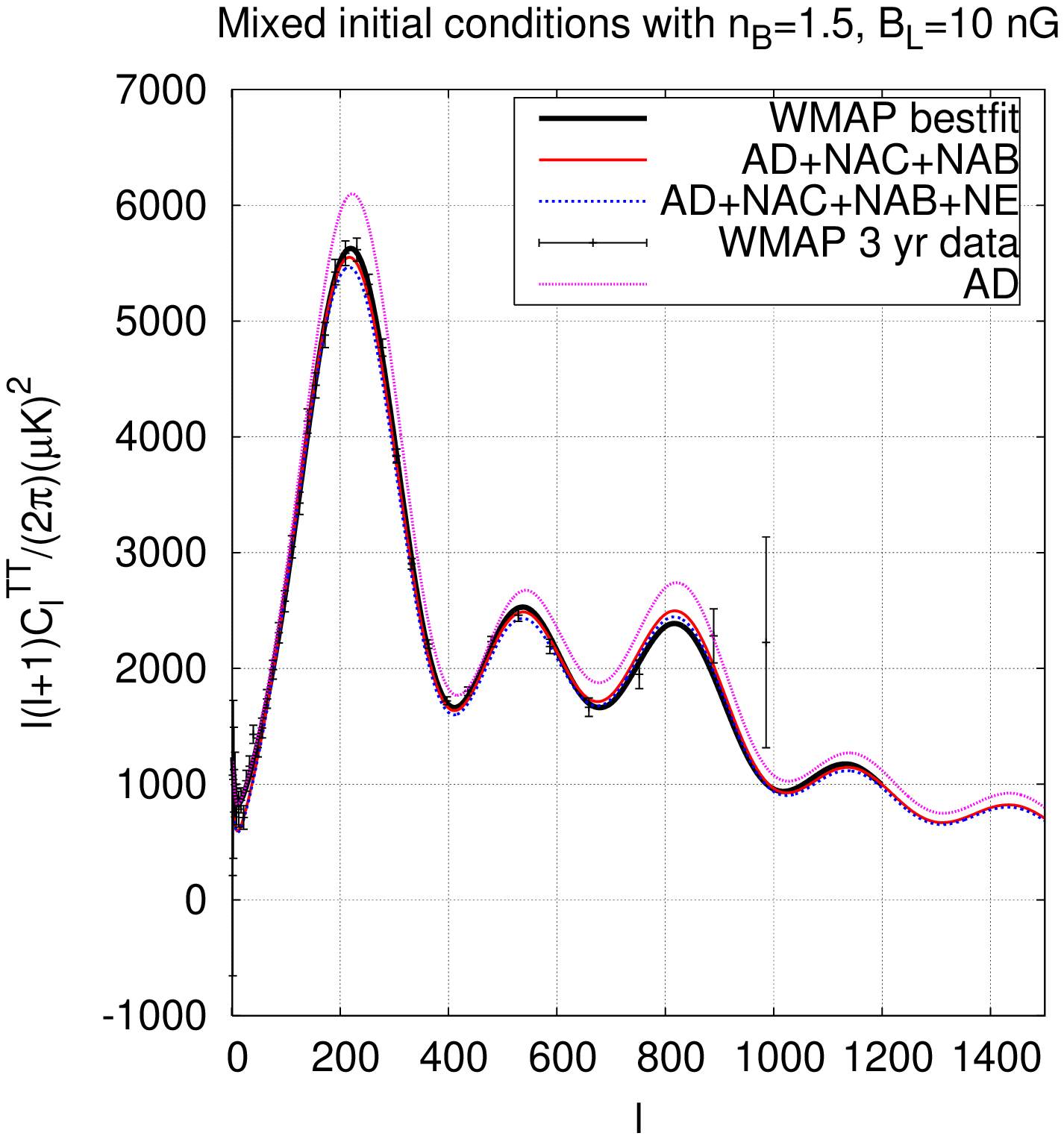}
\includegraphics[height=7cm]{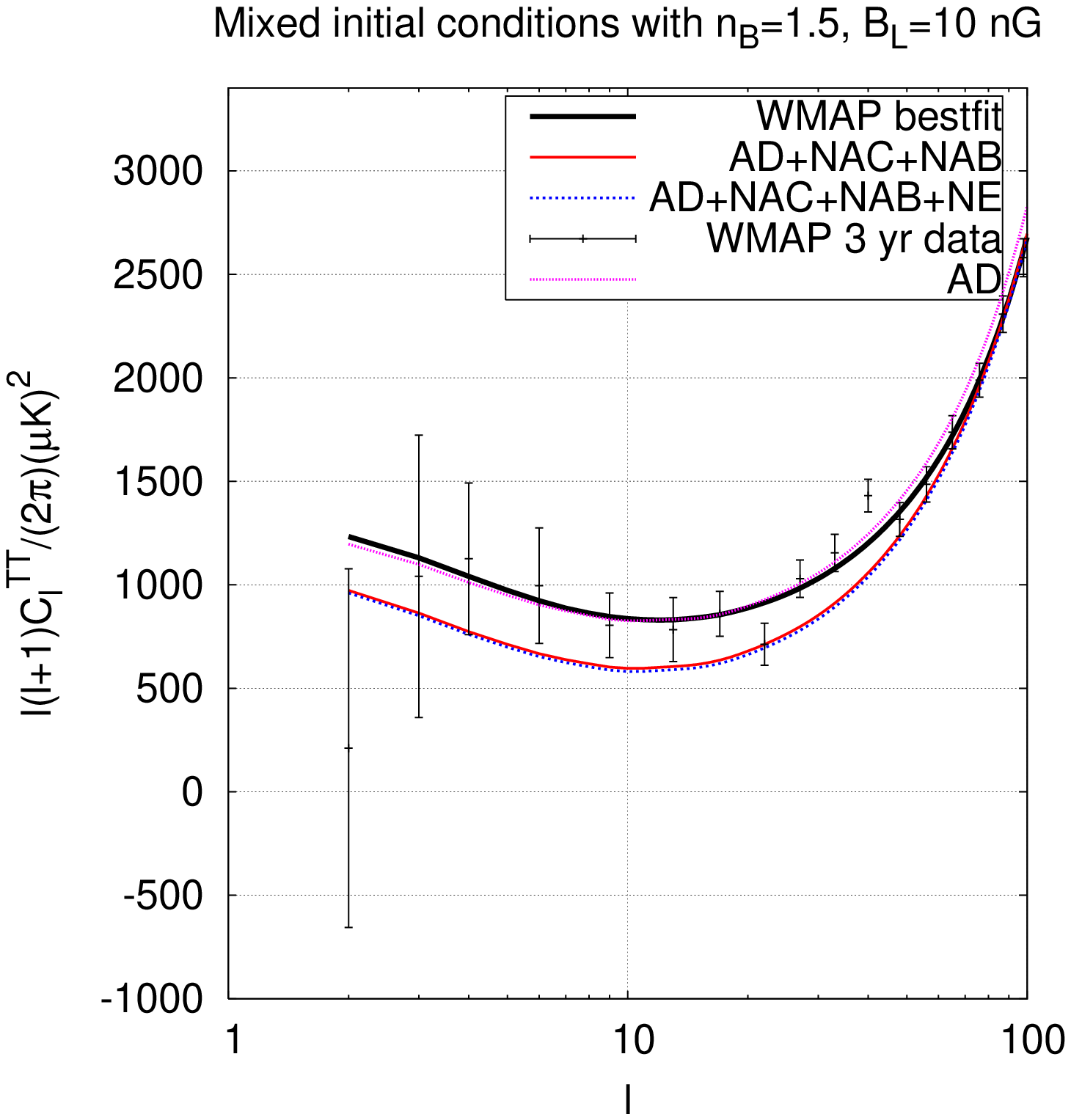}
\includegraphics[height=7cm]{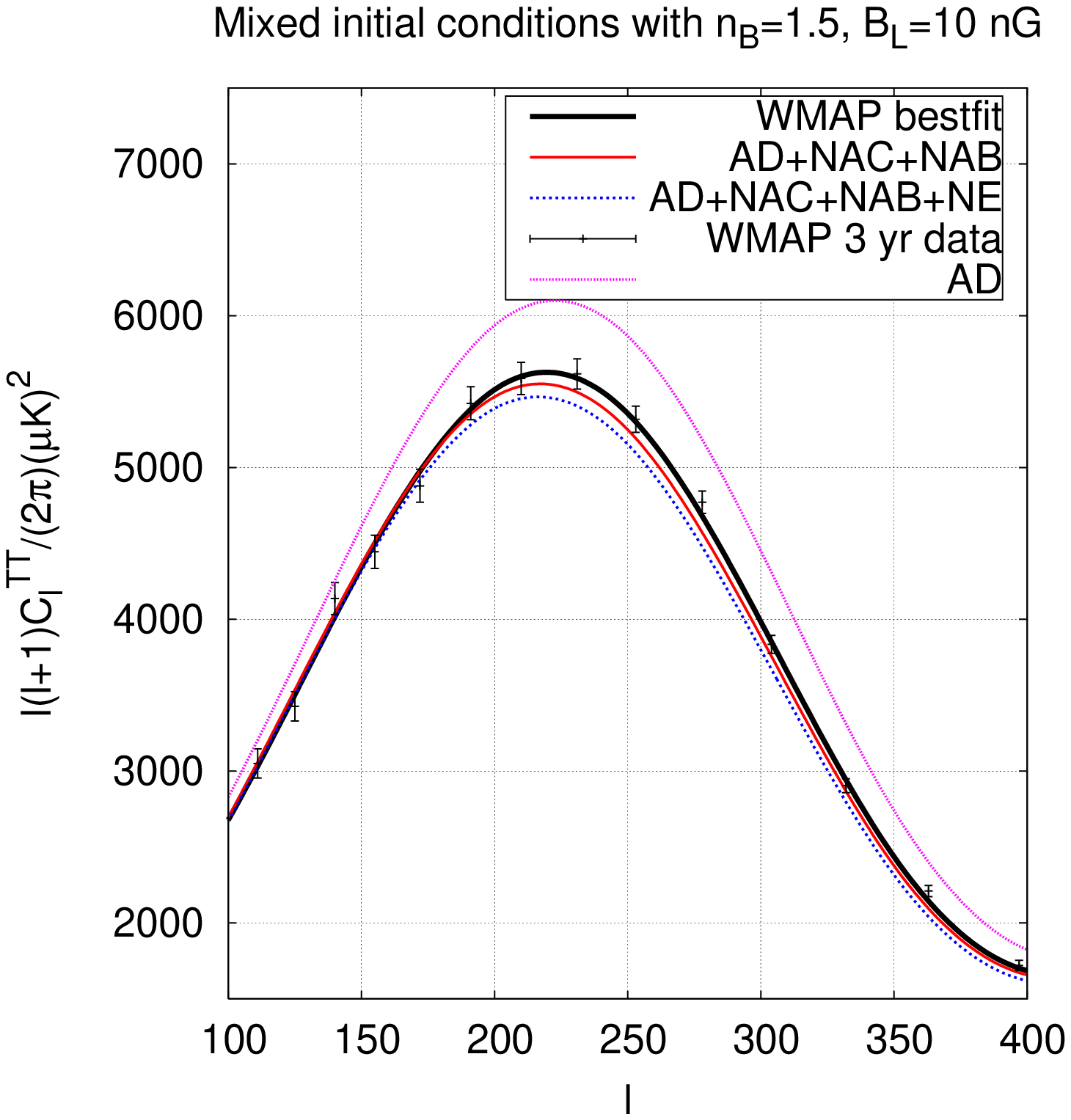}
\includegraphics[height=7cm]{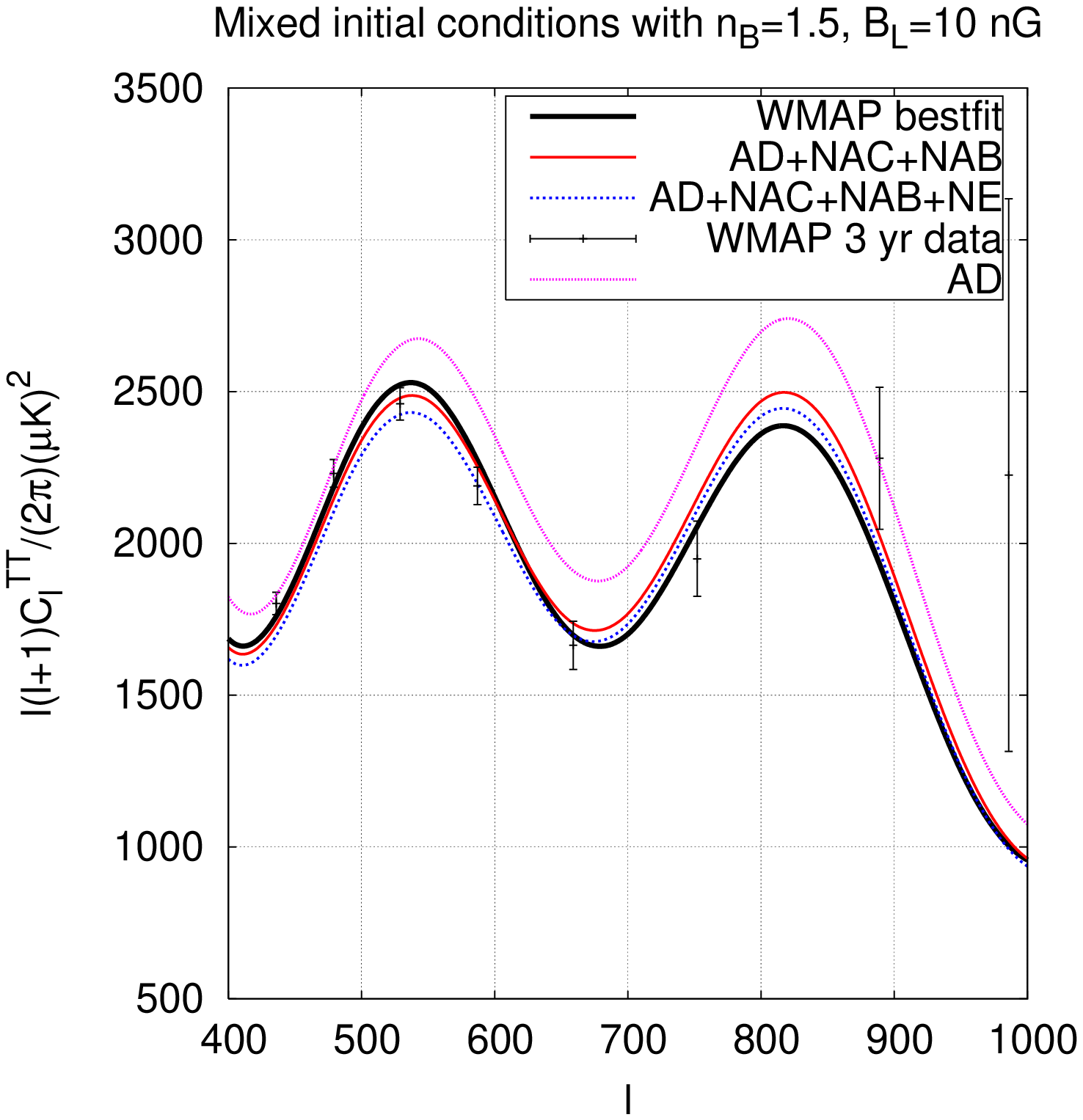}
\caption[a]{The TT angular power spectra in three different cases: AD 
(magnetized adiabatic mode), AD + NAC +NAB (mixture of the magnetized adiabatic mode supplemented by the magnetized CDM-radiation and baryon-radiation modes), AD + NAC + NAB +NE (same as in the previous 
case but with the addition of a neutrino-entropy component). The spectral 
parameters are fixed as in Eqs. (\ref{two}), (\ref{three}) and (\ref{four}).}
\label{Figure5}      
\end{figure}
Generally speaking, 
the relative amplitudes of the different contributions must be 
gauged in such a way that:
\begin{itemize}
\item{} the acoustic peaks are correctly reproduced in the TT angular 
power spectrum;
\item{} the anticorrelation peak in the TE spectra is not 
erased by the addition of the non-adiabatic component;
\item{} the EE angular power spectra should not be too different, at 
large multipoles, from the extrapolated three year best fit to the WMAP 
data alone.
\end{itemize}
The third requirement can be phrased in terms of any best 
fit (also including other data sets, e.g. \cite{LSS1,LSS2} and \cite{SN1,SN2}). This will not change our 
conclusions since, as  far as CMB observables are concerned, 
different best fits lead to central values that are not crucially different for 
the present (illustrative) purposes.
In Fig. \ref{Figure5} the TT correlations are reported for 
different sets of initial conditions. In the top-left plot the 
full range of multipoles is discussed and in the remaining three 
plots the low, intermediate and large $\ell$ regions 
are specifically scrutinized. As the legends specify the full thick line 
denoted the three year best fit to the WMAP data 
alone. 
The cosmological parameters (i.e. $\omega_{\mathrm{b0}}$, $\omega_{\mathrm{c0}}$, $n_{\mathrm{s}}$, $\tau$ and ${\mathcal A}_{{\mathcal R}}$) are hence fixed to the values already assumed for the 
previous plots.  The light curve denotes the magnetized 
adiabatic mode in the case $B_{\mathrm{L}} = 10$ nG and $n_{\mathrm{B}}= 1.5$. Inspection 
of the plots reported in Fig. \ref{Figure5} 
suggests that the case labeled by AD (i.e. a single magnetized 
adiabatic mode) is incompatible with the experimental data.
This (expected) conclusion can be  partially evaded if the magnetized 
adiabatic mode is combined with a mixture of initial conditions.
Two different cases will then be considered:
\begin{itemize}
\item{} AD + NAC+NAB denotes the case where, on top 
of the magnetized adiabatic mode (i.e. AD), the initial conditions include also two other non-adiabatic contributions
which are chosen to be the magnetized CDM-radiation (i.e. NAC) and the magnetized baryon-radiation (i.e. NAB) modes derived, respectively, in Eqs. (\ref{CDS1})--(\ref{CDS11}) and in Eqs. (\ref{BS1})--(\ref{BS11});
\item{}  AD+ NAC+ NAB+ NE denotes the case where, on top 
of the  three contributions mentioned in the previous point there is also a neutrino-entropy mode 
(i.e. NE) which has been derived in Eqs. (\ref{NS1})--(\ref{NS10}).
\end{itemize}
In the case AD + NAC + NAB the spectral parameters are fixed, according 
to the present notations as:
\begin{equation}
{\mathcal A}_{{\mathcal S}} = 0.1 
{\mathcal A}_{{\mathcal R}},\qquad n_{\mathrm{c}} = 1.5, \qquad {\mathcal A}_{\overline{{\mathcal S}}}= 0.3 {\mathcal A}_{{\mathcal R}},\qquad n_{\mathrm{b}} =1.5.
\label{two}
\end{equation}
For the case AD+ NAC+NAB+NE, the spectral 
parameters of the neutrino-entropy mode are fixed as
\begin{equation}
{\mathcal A}_{\tilde{{\mathcal S}}}= 0.025 {\mathcal A}_{{\mathcal R}},\qquad n_{\nu} = 1.4,
\label{three}
\end{equation}
while the spectral parameters of the other two non-adiabatic modes are chosen as in Eq. (\ref{two}).
For both examples summarized in Eqs. (\ref{two}) and (\ref{three}) the remaining spectral 
parameters are selected as: 
\begin{equation}
B_{\mathrm{L}} = 10\, \mathrm{nG}, \qquad n_{\mathrm{B}} =1.5,\qquad 
{\mathcal A}_{{\mathcal R}} = 2.35 \times 10^{-9},
\label{four}
\end{equation}
The values of  the other cosmological parameters is the 
one suggested by the three year best fit to the WMAP data 
alone.
 \begin{figure}[!ht]
\centering
\includegraphics[height=7cm]{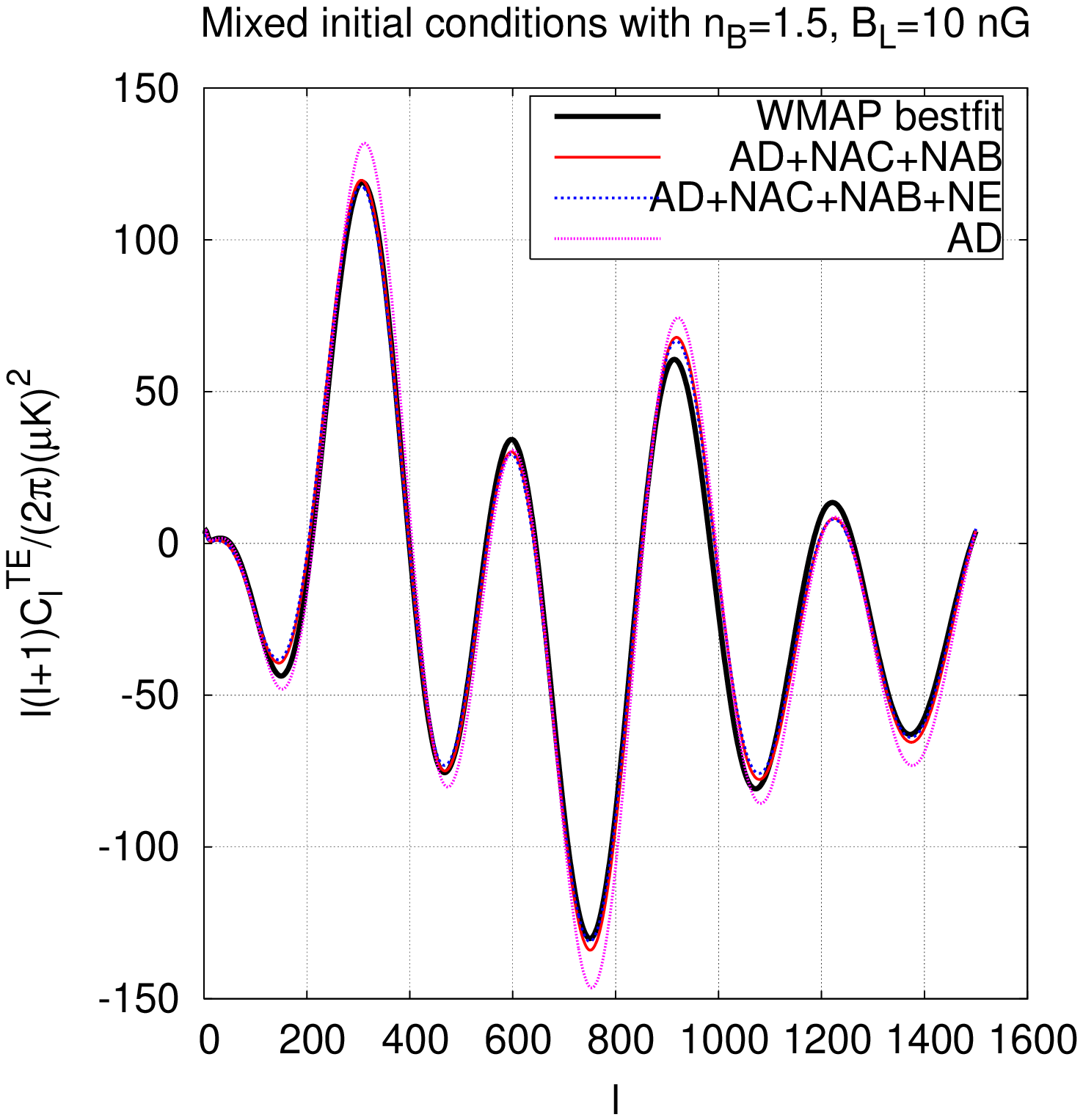}
\includegraphics[height=7cm]{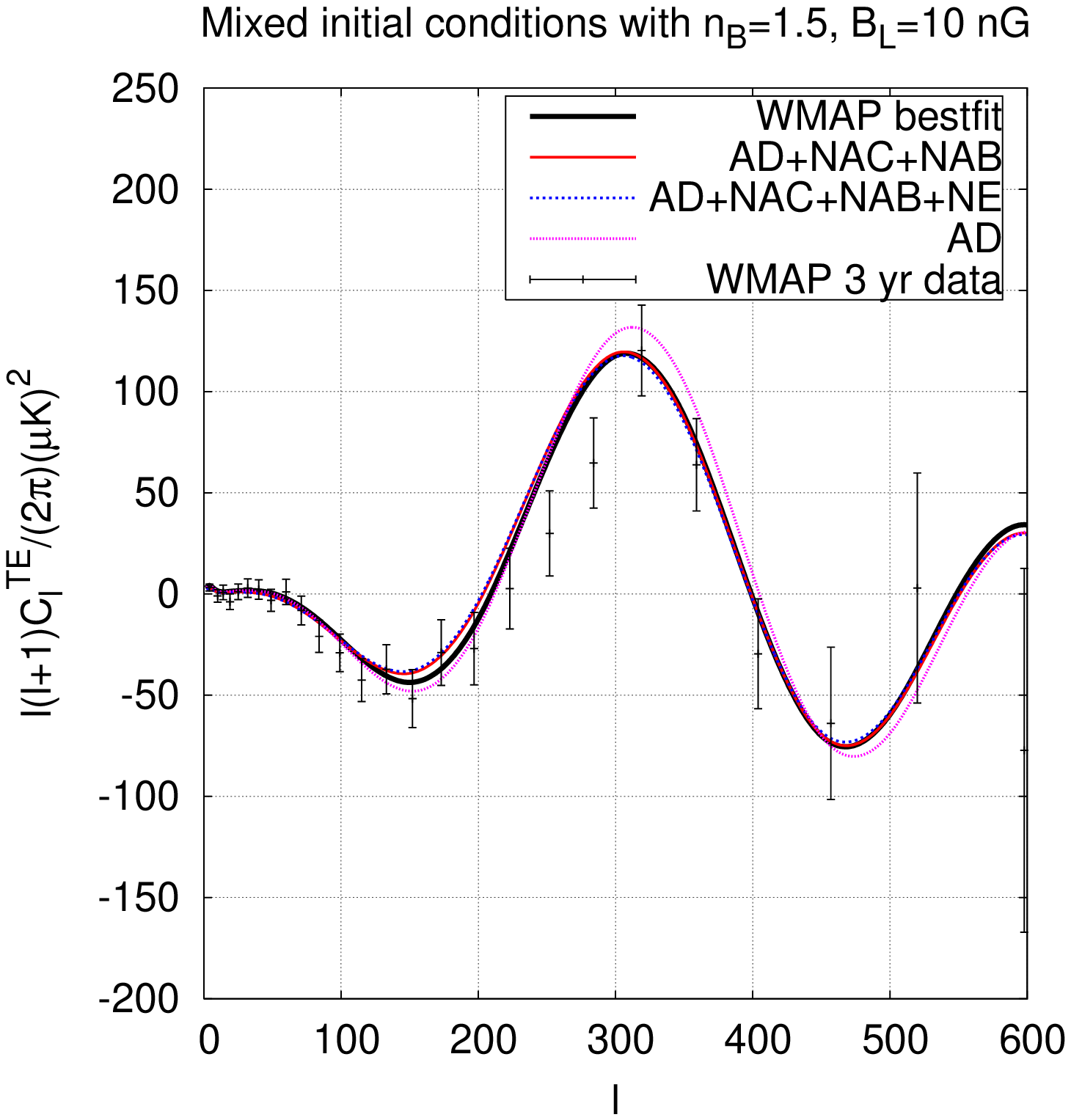}
\includegraphics[height=7cm]{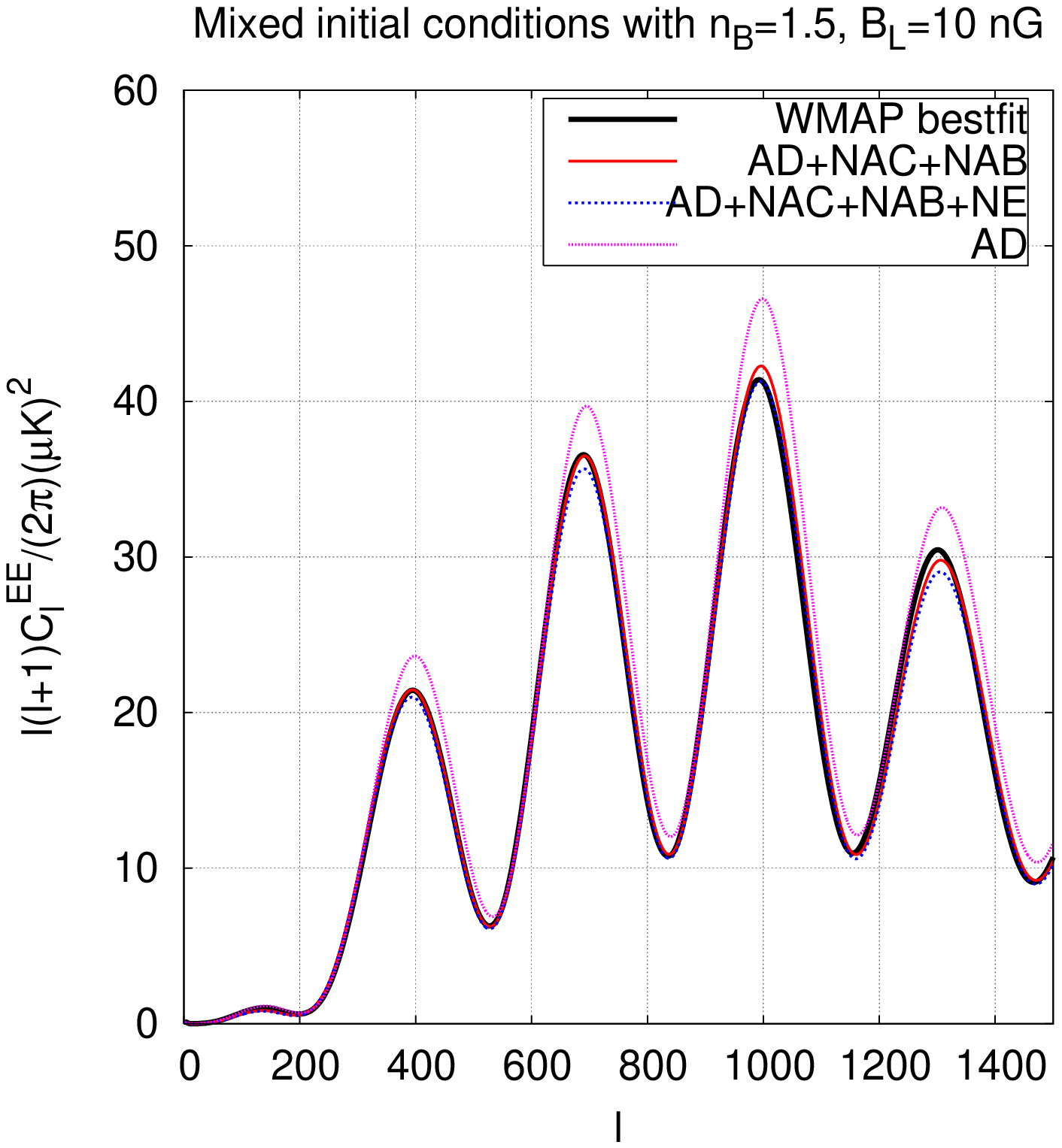}
\includegraphics[height=7cm]{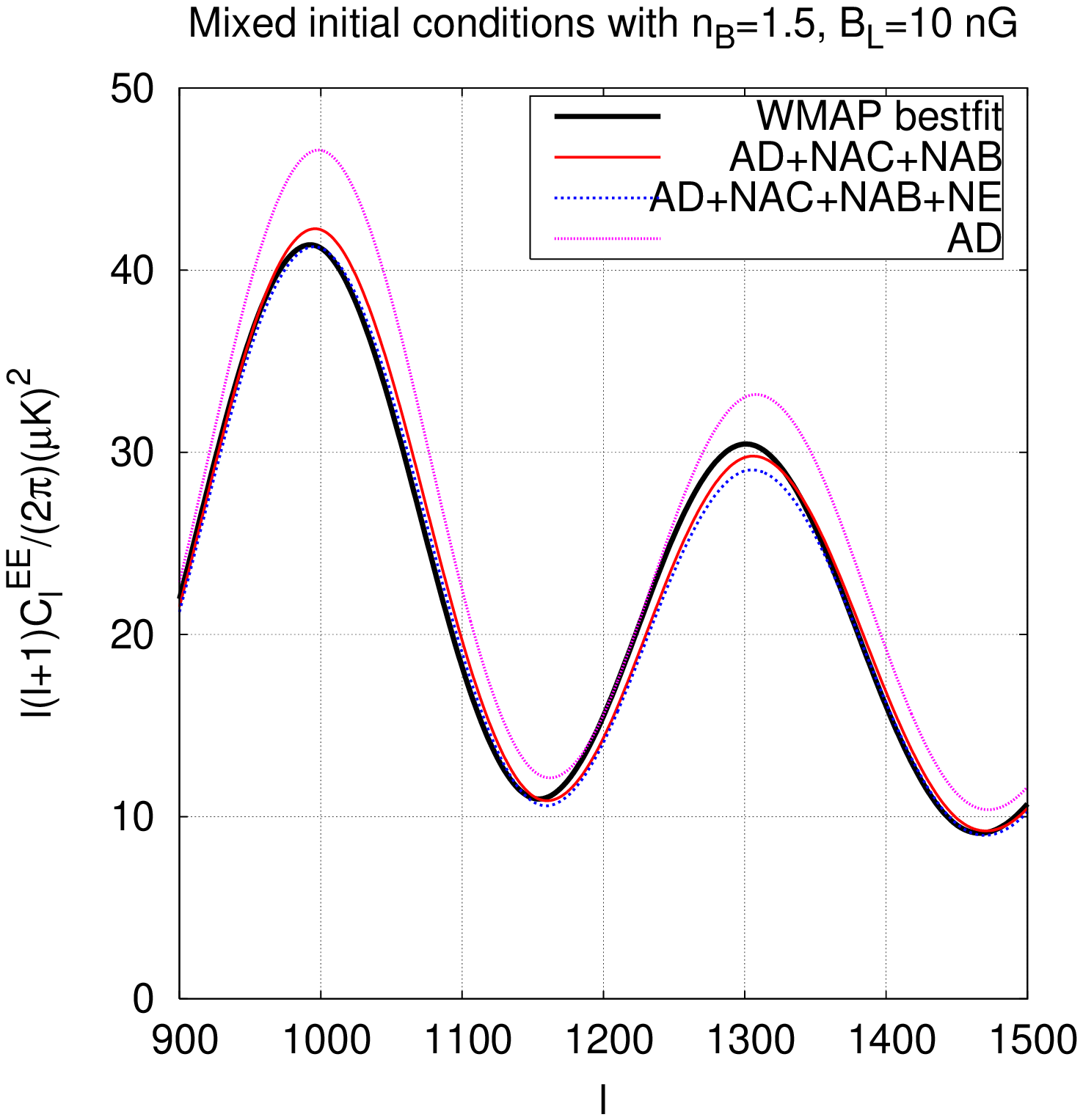}
\caption[a]{The TE and EE angular power spectra are illustrated 
for the same mixtures of initial conditions reported in Fig. \ref{Figure5}.}
\label{Figure6}      
\end{figure}
According to Fig. \ref{Figure5}, the excessive  
increase (and distortion) of the acoustic peaks 
induced by the bare magnetized adiabatic mode (i.e. AD) 
is compensated when the initial conditions are of the type 
AD + NAC + NAB.  The NAC contribution, as discussed 
in Fig. \ref{Figure2} tends to compensate the increase 
of the peaks while the NAB contribution acts on the phases 
(see also Fig. \ref{Figure3}).  As it is 
clear from the bottom-right plot of Fig. \ref{Figure5}
the AD+NAC+NAB combination is even too effective 
in lowering the peaks. The inclusion of the neutrino-entropy 
mode moderates this trend by adding extra-power in  the acoustic region.
This effect arises since the neutrino-entropy mode has a different phase structure 
in comparison with the non-adiabatic mode of the CDM and baryon sector.
From the different ranges of multipoles where the TT power spectra 
have been computed two opposite remarks seem to emerge:
\begin{itemize}
\item{} on the one hand the mixing of different magnetized modes in 
the pre-equality initial conditions seems to shade 
the effect of a (pretty strong) magnetic field;
\item{} on the other hand the reported WMAP error bars 
also suggest that the different cases can still be distinguished.
\end{itemize}
In the light of these remarks it is useful to consider 
the plots reported in Fig. \ref{Figure6} where 
the TE and EE angular power spectra are illustrated. 
The sub-dominant non adiabatic components imply 
that the leading effect in the TE anticorrelation peak 
is determined by the magnetized adiabatic mode. At the same time
the sharp increase of the peaks in the TE and EE spectra 
(for high multipoles) is partially compensated by the non-adiabatic 
components.  For $\ell > 1000$ the distortions induced by the magnetized 
adiabatic mode can be reduced but not eliminated.

The reported examples are, in our opinion, rather effective for motivating the approach 
pursued in the present research program which is directly linked to the possibilities 
of achieving accurate determinations of the temperature and polarization observables 
with the forthcoming Planck explorer mission \cite{planck}.
A systematic approach in this direction is 
a necessary step for any quantitative strategy 
of parameter extraction involving 
large-scale magnetic fields.
\renewcommand{\theequation}{8.\arabic{equation}}
\setcounter{equation}{0}
\section{Concluding discussion}
\label{sec8}
More than fifty years ago large-scale magnetic fields entered the scene 
because of the interplay with cosmic ray physics \cite{fermi}. Today cosmic ray physics 
is again confronted with the relevance of large-scale magnetic fields 
to infer more fundamental conclusions on the nature of the observed fluxes 
of ultra-high energy cosmic rays \cite{auger}. In the intervening lapse of time 
radio-astronomy has established that the largest gravitationally bound systems 
are all magnetized to various degrees. Still the question remains: where do the 
observed fields come from? 

The standard paradigm used to interpret the cosmological data, i.e. the $\Lambda$CDM model, 
does not contemplate the presence of a magnetized contribution very much in the same 
way as, in the pivotal $\Lambda$CDM mode tensors are absent. 
The increase in the quality of astrophysical observations will allow, in the near future, to 
include further parameters to the basic $\Lambda$CDM scenario. In \cite{giokunz1}
we formulated a plausible extension of the $\Lambda$CDM scenario, i.e the magnetized 
$\Lambda$CDM model (m$\Lambda$CDM). In the minimal m$\Lambda$CDM model 
the CMB initial conditions are simply dictated by the magnetized 
adiabatic mode where no entropic modes are allowed. 
In the present paper the minimal m$\Lambda$CDM model has been 
extended to the case of magnetized initial conditions with entropic components.
The spirit of the present investigation is, at once,  modest and conservative: we just want to bring the treatment 
of magnetized CMB anisotropies to the same standards which are typical of the cases 
where large-scale magnetic fields are absent. The problem with this 
plan is that specific numerical tool have to be developed. Unlike 
more standard additions to the $\Lambda$CDM paradigm (which are customarily included 
in the standard codes) large-scale magnetic fields demand dedicated numerical 
approaches which, so far, did not exist. The proof of this statement is that, before 
the present analysis, we did not know, for instance, the effects of large-scale 
magnetic fields on the various non-adiabatic modes contemplated in the standard 
CMB initial conditions. 

In the present paper the initial conditions of magnetized CMB anisotropies 
have been derived in a generalized perspective. The analysis has been conducted both analytically 
on the truncated form of the Einstein-Boltzmann hierarchy.
The CMB observables have then been numerically computed
for each single set of initial conditions to emphasize the relevant shape effects. 
The addition of the magnetic field is not able to 
change the sign of the first peak of the TE correlations so that we can safely conclude 
that a single magnetized non-adiabatic mode is not compatible with the experimental 
observations. The same sharp statement cannot be made, however, if the
initial conditions of the truncated Einstein-Boltzmann hierarchy 
are given in terms of a superposition of modes where the dominant 
amplitude is provided by the magnetized adiabatic mode.
The  examples provided in this paper suggest 
that a thorough analysis of the TT, TE and EE angular power spectra 
allow, in principle, to distinguish the signatures of different magnetized 
initial conditions.  In the near future we plan to pursue our program 
by  sharpening those theoretical tools that 
 might allow for  a direct observational test of the effects 
 of large-scale magnetic fields on CMB anisotropies. While we are aware 
 of the theoretical challenges associated with this effort, the potential relevance 
of the whole strategy justifies our endeavors.

\section*{Acknowledgments}

K. E. Kunze  acknowledges the support of the  ``Ram\'on y Cajal''  program as well as the grants FPA2005-04823, FIS2006-05319 and 
CSD2007-00042 of the Spanish Science Ministry.

\newpage
\begin{appendix}
\renewcommand{\theequation}{A.\arabic{equation}}
\setcounter{equation}{0}
\section{Truncated Einstein-Boltzmann hierarchies}
\label{APPA}
The evolution equations of the truncated Einstein-Boltzmann hierarchy will be illustrated
in the tightly coupled regime which is the relevant one to set initial conditions. For
reasons of convenience the language of the synchronous gauge will be adopted.
Similar equations discussed either in different gauges or in dedicated gauge-invariant 
formalisms can be usefully found elsewhere \cite{mg1,mg2,mg3}. The synchronous 
coordinate system is defined from the perturbed form of the metric tensor in terms 
of two variables denoted, in Fourier space, as $h(k,\tau)$ and $\xi(k,\tau)$:
\begin{equation}
\delta_{\mathrm{s}} g_{ij}(k,\tau) = a^2 \biggl[ \hat{k}_{i} \hat{k}_{j} h(k,\tau) + 6\biggl( \hat{k}_{i} \hat{k}_{j} - \frac{\delta_{ij}}{3}\biggr)\xi(k,\tau)\biggr].
\label{metric}
\end{equation}

The bulk velocity of the plasma \footnote{
Within the notations of the present paper $\theta_{X} = \vec{\nabla}\cdot \vec{v}_{X}$ is the divergence 
of the peculiar velocity of a given species labeled by $X$. In the text we will simply talk about 
"velocity field" but it is understood that this terminology denotes the three-divergence of the velocity 
field.}
$\theta_{\mathrm{b}}$ coincides with the velocity of the baryons which are 
coupled to the electrons through Coulomb scattering. Also the photons are coupled 
to the baryons because of Thompson scattering. Deep in the radiation-dominated epoch, when 
the initial conditions of the Boltzmann hierarchy are ideally set, the baryon velocity field $\theta_{\mathrm{b}}$ and the photon velocity field (i.e. $\theta_{\gamma}$) are very well synchronized because 
the Thompson rate is much larger, at the corresponding epoch, than the Hubble 
rate. To zeroth-order in the tight-coupling expansion, therefore, $\theta_{\gamma} \simeq \theta_{\mathrm{b}} = 
\theta_{\gamma\mathrm{b}}$ where $\theta_{\gamma\mathrm{b}}$ denotes the common value 
of the baryon-photon velocity. At later times, well after equality, the primeval plasma 
recombine and the baryon velocity will no longer equal the photon velocity. This 
effect is taken into account in the numerical integration, however, at early time, 
the zeroth-order in the tight-coupling limit is an excellent approximation and will be 
the one used to set, consistently the initial conditions of the Einstein-Boltzmann 
hierarchy. Higher-orders in the tight-coupling expansion can be used 
also for semi-analytical estimates \cite{mg1,mg2}.

In the synchronous gauge the perturbed Einstein equations for the system 
of magnetized perturbations read \cite{giokunz2}:
\begin{eqnarray}
&& 2 k^2 \xi - {\mathcal H} h' = 8\pi G a^2 ( \delta_{\mathrm{s}} \rho_{\mathrm{t}} + \delta_{\mathrm{s}} \rho_{\mathrm{B}}),
\label{HAM}\\
&& k^2 \xi' = - 4 \pi G a^2 (p_{\mathrm{t}} + \rho_{\mathrm{t}}) \theta_{\mathrm{t}},
\label{MOM}\\
&& h'' + 2 {\mathcal H} h' - 2 k^2 \xi = 24\pi G a^2 (\delta_{\mathrm{s}} p_{\mathrm{t}} + \delta_{\mathrm{s}} p_{\mathrm{B}}),
\label{heq}\\
&& ( h + 6 \xi)'' + 2 {\mathcal H} ( h + 6 \xi)' - 2 k^2 \xi = 24 
\pi G a^2 [(p_{\nu} + \rho_{\nu})\sigma_{\nu} + (p_{\gamma} + 
\rho_{\gamma}) \sigma_{\mathrm{B}}],
\label{anisstr}
\end{eqnarray}
where the following global variables have been defined:
\begin{equation}
(p_{\mathrm{t}} + \rho_{\mathrm{t}}) \theta_{\mathrm{t}} = \sum_{\mathrm{a}} (p_{\mathrm{a}} + \rho_{\mathrm{a}}) \theta_{\mathrm{a}},\qquad \delta_{\mathrm{s}} \rho_{\mathrm{t}} = \sum_{\mathrm{a}} \delta_{\mathrm{s}} \rho_{\mathrm{a}}, \qquad
\delta_{\mathrm{s}} p_{\mathrm{t}} = \sum_{\mathrm{a}} \delta_{\mathrm{s}} p_{\mathrm{a}} = w_{\mathrm{a}} \delta_{\mathrm{s}} \rho_{\mathrm{a}}.
\label{global}
\end{equation}
The sums appearing in Eq. (\ref{global}) extends over the four species of the plasma (i.e. photons, neutrinos, 
baryons and CDM particles) and $w_{\mathrm{a}}$ is the barotropic index of each species.
Equations (\ref{HAM}) and (\ref{MOM}) are, respectively, the Hamiltonian and momentum constraints coming from the $(00)$ and $(0i)$ components of the perturbed Einstein equations.
By using the Friedmann-Lema\^itre equations in the form:
\begin{equation}
3 {\mathcal H}^2 = 8\pi G a^2 \rho_{\mathrm{t}},\qquad {\mathcal H}^2 -{\mathcal H}' = 4 \pi G a^2 (p_{\mathrm{t}}  + \rho_{\mathrm{t}})
\label{FL}
\end{equation}
Eqs. (\ref{HAM}) and  (\ref{MOM}) can also be written in more explicit terms as 
\begin{eqnarray}
&&2 k^2 \xi - {\mathcal H} h' = 3 {\mathcal H}^2\biggl\{ \Omega_{\mathrm{R}} (R_{\gamma} \delta_{\gamma} + 
R_{\nu} \delta_{\nu}) + R_{\gamma} \Omega_{\mathrm{R}} \Omega_{\mathrm{B}}  + 
 \Omega_{\mathrm{M}} \biggl[\biggl(\frac{\omega_{\mathrm{c}0}}{\omega_{\mathrm{M}0}}\biggr) \delta_{\mathrm{c}}
+ \biggl(\frac{\omega_{\mathrm{b}0}}{\omega_{\mathrm{M}0}}\biggr)\delta_{\mathrm{b}}\biggr]\biggr\},
\label{HAM1}\\
&& k^2 \xi' = - 2 {\mathcal H}^2 \biggl[ \Omega_{\mathrm{R}} R_{\nu}\theta_{\nu} +  \Omega_{\mathrm{R}} R_{\gamma}( 1 + R_{\mathrm{b}}) \theta_{\gamma\mathrm{b}} + \frac{3}{4} \Omega_{\mathrm{M}} \biggl(\frac{\omega_{\mathrm{c}0}}{\omega_{\mathrm{M}0}}\biggr)\rho_{\mathrm{c}} \theta_{\mathrm{c}} \biggr].
\label{MOM1}
\end{eqnarray}
With analog algebra we will have that Eqs. (\ref{heq}) and (\ref{anisstr}) can be recast in the 
following form:
\begin{eqnarray}
&&  h'' + 2 {\mathcal H} h' - 2 k^2 \xi = 3 {\mathcal H}^2 \Omega_{\mathrm{R}} ( R_{\gamma} \delta_{\gamma} + R_{\nu} \delta_{\nu} +  R_{\gamma} \Omega_{\mathrm{B}})
\label{heq2}\\
&& ( h + 6 \xi)'' + 2 {\mathcal H} ( h + 6 \xi)' - 2 k^2 \xi = 12 {\mathcal H}^2 \Omega_{\mathrm{R}} [ R_{\nu} \sigma_{\nu} + R_{\gamma} \sigma_{\mathrm{B}}],
\label{anisstr2}
\end{eqnarray}
where 
\begin{equation}
\Omega_{\mathrm{R}} = \frac{1}{1 + \alpha}, \qquad \Omega_{\mathrm{M}} = \frac{\alpha}{\alpha + 1}.
\label{OMdef}
\end{equation}
Note that, in the momentum constraint (i.e. Eq. (\ref{MOM1})) we assumed already the tight-coupling 
regime insofar as $\theta_{\gamma} \simeq \theta_{\mathrm{b}} = \theta_{\gamma\mathrm{b}}$.
To solve consistently for the initial conditions of the Einstein-Boltzmann hierarchy we have also 
to specify the evolution equations for the monopoles and the dipoles of the various species. The 
evolution equations of the monopoles can be written, respectively, as:
\begin{eqnarray}
&& \delta_{\nu}' = - \frac{4}{3} \theta_{\nu} + \frac{2}{3} h',
\label{deltanu}\\
&& \delta_{\gamma}' = - \frac{4}{3} \theta_{\gamma\mathrm{b}} + \frac{2}{3} h',
\label{deltagamma}\\
&& \delta_{\mathrm{b}}' = -\theta_{\gamma\mathrm{b}} + \frac{h'}{2},
\label{deltab}\\
&& \delta_{\mathrm{c}}' = - \theta_{\mathrm{c}} + \frac{h'}{2}.
\label{deltac}
\end{eqnarray}
The evolution equations for the dipoles are given, instead, as 
\begin{eqnarray}
&&\theta_{\nu}' = \frac{k^2}{4} \delta_{\nu} - k^2 \sigma_{\nu},
\label{thetanu}\\
&&\theta_{\gamma\mathrm{b}}'+ \frac{{\mathcal H} R_{\mathrm{b}}}{R_{\mathrm{b}} +1 } \theta_{\gamma\mathrm{b}} = \frac{k^2}{4 ( 1 + R_{\mathrm{b}})} \delta_{\gamma} + \frac{k^2}{4 ( 1 + R_{\mathrm{b}})} (\Omega_{\mathrm{B}} - 4 \sigma_{\mathrm{B}}),
\label{thetab}\\
&& \theta_{\mathrm{c}} + {\mathcal H} \theta_{\mathrm{c}} =0.
\label{thetac}
\end{eqnarray}
In Eq. (\ref{thetanu}) there appears also $\sigma_{\nu}$ and, consequently, the corresponding 
evolution equation is given by:
\begin{equation}
\sigma_{\nu}' = \frac{4}{15} \theta_{\nu} - \frac{3}{10} k {\mathcal F}_{\nu 3} - \frac{2}{15} h' - \frac{4}{5} \xi',
\label{sigmanu}
\end{equation}
where ${\mathcal F}_{\nu 3}$ is the octupole of the neutrino phase-space distribution.

Recalling that the fluctuation of the total pressure can be expressed as the sum of 
the adiabatic contribution and the non-adiabatic contribution (see Eq. (\ref{deltap})), the generalized 
evolution equation for the curvature perturbations can be obtained. This equation 
is used, in Section \ref{sec2} to get the first iteration of the large-scale 
solution which will allow to determine the various modes. 
By taking the difference between Eqs. (\ref{heq}) and (\ref{anisstr}) we do get that 
\begin{equation}
\xi'' + 2 {\mathcal H} \xi' = 4\pi Ga^2[ (p_{\nu} + \rho_{\nu}) \sigma_{\nu} + (p_{\gamma} + \rho_{\gamma})\sigma_{\mathrm{B}} - \delta_{\mathrm{s}} p_{\mathrm{t}} - \delta_{\mathrm{s}} p_{\mathrm{B}}].
\label{der1}
\end{equation}
By now summing up Eq. (\ref{der1}) and Eq. (\ref{HAM1}) (appropriately multiplied by $c_{\mathrm{st}}^2$)
the following equation can be obtained:
\begin{eqnarray}
&&\xi'' + {\mathcal H}( 2 + 3 c_{\mathrm{st}}^2)\xi' = \frac{{\mathcal H}}{2} c_{\mathrm{st}}^2 (h' + 6 \xi') - k^2 
c_{\mathrm{st}}^2 \xi 
\nonumber\\
&&+ 4\pi G a^2 \biggl[ (p_{\nu} + \rho_{\nu}) \sigma_{\nu} + (p_{\gamma} + \rho_{\gamma}) \sigma_{\mathrm{B}}+  \biggl( c_{\mathrm{st}}^2 - \frac{1}{3}\biggr) \delta_{\mathrm{s}} \rho_{\mathrm{B}}
- \delta_{\mathrm{s}} p_{\mathrm{nad}}\biggr].
\label{der2}
\end{eqnarray}
But now recall that the curvature perturbations in the synchronous gauge simply become:
\begin{equation}
{\mathcal R} = \xi + \frac{{\mathcal H}\xi'}{{\mathcal H}^2 - {\mathcal H}'},\qquad 
{\mathcal R}' = \frac{2}{3} \frac{\rho_{\mathrm{t}}}{\rho_{\mathrm{t}} + p_{\mathrm{t}}} 
\biggl[ \frac{\xi''}{{\mathcal H}} + (2 + 3 c_{\mathrm{st}}^2)\xi'\biggr],
\label{der3}
\end{equation}
where the second relation can be obtained from the first time derivative of ${\mathcal R}$ since, by definition, 
$c_{\mathrm{st}}^2 = p_{\mathrm{t}}'/\rho_{\mathrm{t}}'$. Finally, combining Eqs. (\ref{der2}) and (\ref{der3}),
we do get the wanted equation, i.e. 
\begin{eqnarray}
&& {\mathcal R}' = - \frac{{\mathcal H}}{p_{\mathrm{t}} + \rho_{\mathrm{t}}} \delta_{\mathrm{s}} p_{\mathrm{nad}} + 
\frac{{\mathcal H}}{p_{\mathrm{t}} + \rho_{\mathrm{t}}} \biggl( c_{\mathrm{st}}^2 - \frac{1}{3}\biggr) \delta_{\mathrm{s}} \rho_{\mathrm{B}}
\nonumber\\
&& + \frac{{\mathcal H}^2 (h' + 6 \xi')}{8\pi G a^2 ( p_{\mathrm{t}} + \rho_{\mathrm{t}})} - \frac{{\mathcal H} k^2 c_{\mathrm{st}}^2 \xi}{4\pi G a^2 (p_{\mathrm{t}} + \rho_{\mathrm{t}})} + 
\frac{{\mathcal H}[(p_{\nu} + \rho_{\nu})\sigma_{\nu} + (p_{\gamma} + \rho_{\gamma})\sigma_{\mathrm{B}}]}{(p_{\mathrm{t}} + \rho_{\mathrm{t}})} .
\label{der4}
\end{eqnarray}
The terms appearing in the second line of Eq. (\ref{der4}) are of higher order in $k/{\mathcal H}$ while 
the terms appearing in the first line are responsible for the leading order solution in the limit $k\tau \ll 1$.
Equation (\ref{der4}) will be used in Section \ref{sec2} to obtain the the evolution of the various modes 
in the limit of wavelengths larger than the Hubble radius.
\end{appendix}

\end{document}